\documentstyle[preprint,prb,aps]{revtex}
\begin{document}
\draft
\title
{Kondo lattice model: Unitary transformations, spin dynamics,\\
strongly correlated charged modes, and vacuum instability.}
\author
{J.M. Prats and F. L\'opez-Aguilar}
\address
{Departament de F\'\i sica, Grup d'Electromagnetisme,
Universitat Aut\`onoma de Barcelona,\\
Bellaterra, E-08193 Barcelona, Spain}
\maketitle
\subsection*{Abstract}

Using unitary transformations, we express the Kondo lattice 
Hamiltonian in terms of fermionic operators that annihilate the 
ground state of the interacting system and that represent the best 
possible approximations to the actual charged excitations. In this 
way, we obtain an effective Hamiltonian which, for small couplings, 
consists in a kinetic term for conduction electrons and holes, an
RKKY-like term, and a renormalized Kondo interaction. The physical 
picture of the system implied by this formalism is that of a vacuum 
state consisting in a background of RKKY-induced spin correlations,
where two kinds of elementary modes can be excited: Soft neutral modes 
associated with deformations of the spin liquid, which lead to
very large low-temperature values of the heat capacity and magnetic
susceptibility, and charged modes corresponding to the excitation 
of electrons and holes in the system. Using the translational and spin 
rotational symmetries, we construct a simple ansatz to determine the 
charged excitations neglecting the effects of the spin correlations.
Apart from the `normal', uncorrelated states, we find strongly correlated
charged modes involving soft electrons (or holes) and spin fluctuations,
which strongly renormalize the low-energy charged spectrum, and whose
energy becomes negative beyond a critical coupling, signaling a vacuum 
instability and a transition to a new phase.

\

\noindent PACS: 71.27.+a, 75.30.Mb, 75.20.Hr.

\vfill
\newpage

\section {Introduction}

Despite many years of intense experimental and theoretical studies 
of heavy-fermion (HF) systems,\cite{fisk,stewart,lee,fulde} 
their physical picture still remains 
controversial.  It is generally believed that both HF metals and the 
so-called mixed-valence compounds are well-represented by the 
Anderson lattice model (ALM) and that their phenomenologies 
strongly depend on the location of the f band with respect to the 
Fermi level. Ce-based HF compounds are usually regarded as an extreme 
case in which the deep f level and high on-site Coulomb repulsion 
strongly inhibit f-charge fluctuations. This has led many authors
to regard the Kondo lattice model (KLM) as a canonical model for
these compounds since it has been proved to be the limit of the 
ALM for vanishing f-charge fluctuations,\cite{schrieffer} 
and it exclusively involves 
the degrees of freedom that are expected to be relevant to the 
low-energy phenomenology of these materials.

Among the most important features characterizing this phenomenology
are the very large values of the low-temperature magnetic
susceptibility and linear coefficient of the specific heat,
the appearance, in some cases, of antiferromagnetic order with
strongly quenched local moments, and the transition to
a superconducting phase in systems like CeCu${}_2$Si${}_2$.

HF metals are often regarded as `dense Kondo systems' and its properties
qualitatively discussed extending to the lattice the results obtained 
for the single\cite{mahan,fulde2} and two-impurity\cite{jones}
cases. In the first of these cases
(Kondo problem), a singlet state is formed at zero temperature between
the spins of the impurity and of the conduction electrons.\cite{yosida}
Since the low-energy singlet-triplet excitation associated with this
(Abrikosov-Shul) resonance gives a 
contribution to the specific heat,\cite{wilkins}
the formation of one of such resonances at each site of a Kondo lattice
would account for the very high heat capacities measured in these 
systems. In the two-impurity case, however, 
we have an additional ingredient,
which is the appearance of an induced, inter-site (RKKY) interaction.
Thus, magnetic order in the Kondo lattice is often interpreted as
due to a competition between the RKKY interaction (which tends to
produce magnetic order) and the Kondo effect (which tends to quench
the local moments).\cite{doniach}

These interpretations, however, 
have been criticized by Nozi\`eres,\cite{nozieres}
who pointed out that only a very narrow layer of conduction electrons
around the Fermi level are energetically capable of forming Kondo
singlets, which implies that we can have only one `efficient'
electron per {\em many} local spins. Moreover, Nozi\`eres argument
implies that, if strongly correlated (SC) states between conduction
electrons and local moments are formed in the system, they should
not correspond to local Kondo singlets but to extended, collective
states in which one electron couples to {\em many} local moments,
namely, to a spin wave. The calculation of such SC states or
`magnetic polarons' is, in fact, one of the main objectives of this
article.

Another consequence of the Nozi\`eres argument is that only the 
very-low-energy part of the electronic dispersion relation can
be strongly renormalized and, obviously, this mass renormalization 
cannot account for a measured entropy at the characteristic Kondo lattice
temperature ($T_{KL}$) of the order of $N \ln L$, where $L$ and $N$ are 
the f-level degeneracy and number of lattice sites respectively. Since
conduction electrons are the only source of electric charge in the
KLM, charged fermions cannot be responsible for the huge masses
implied by the low-temperature specific heat and magnetic
susceptibility. 

In contrast to the slave-boson\cite{coleman,millis} 
and variational\cite{rice} approximations,
which defend the existence of {\em charged} heavy particles,
the above arguments have led Kagan, Kikoin, and Prokof'ev\cite{kagan}
to suggest that the main contribution to the low-temperature
thermodynamics of these materials comes from {\em neutral} Fermi
excitations of {\em spin origin}. They show\cite{kagan} that
this picture is in good agreement with the available experimental
data while a charged-fermion-based approach seems to be in
conflict with De Haas-Van Alphen experiments.\cite{kagan2}

In this paper, we introduce a mathematical formalism to study
the KLM that leads to a physical picture of HF systems (summarized
in the conlusions) which does not enter in conflict with the 
Nozi\`eres exhaustion problem and which, among other things, supports the 
views presented in Ref. 16.

The basic idea of this formalism is presented in Section II, and
consists in carrying out a unitary transformation to determine
a new set of canonical fermionic operators that annihilate the 
ground state of the interacting system. We find that, when the
Kondo lattice Hamiltonian (KLH) is expressed in terms of these operators,
it essentially consists, for small couplings, in a kinetic term
for electrons and holes containing a gap in its dispersion relation,
an RKKY term, and a Kondo interaction in which the terms involving
an electron and a hole operator have been cancelled.

Using this expression, we discuss magnetic order in Section III.
We argue that, since all the terms of $H$ that contain fermions
annihilate the vacuum state determined in Section II, the RKKY
interaction is the only one responsible for the dynamics of
the spin lattice unless some electronic instability (which we
indeed find for stronger couplings) takes place. Thus, we believe
that the reasons for an eventual formation of a spin liquid should 
be seeked in the enhancement of spin fluctuations due to the strong
frustration produced by the long-range RKKY interaction.
We also argue, in agreement with Ref. 16, that
the very large specific heat measured in these
systems should be attributed to the enormous entropy increase
associated to the thermal breakdown of soft spin correlations.

We mentioned that the dispersion relation corresponding to 
the kinetic term for electrons and holes of the
transformed Hamiltonian contains a gap.
However, if no symmetry is broken and no bound states are formed, 
there is no physical reason for the appearance of a gap in the
middle of the conduction band. In Section IV, we calculate the
{\em actual} (uncorrelated) charged modes neglecting the effect of the
spin correlations and show, to leading order, that no gap is actually
opened at the Fermi level.

As we mentioned when we discussed the Nozi\`eres exhaustion problem,
if SC states are formed in the system, they should be
extended, collective states in which a conduction electron
or hole couples to a spin wave.
In Section V, we look for such states. We argue that, to include
the effect of the high-energy tail of the Kondo interaction, these
modes should be constructed from `optimal' fermionic operators
that represent the best possible approximations to the actual
(uncorrelated) charged modes. These new operators are determined
with the aid of a second unitary transformation that leaves
the ground state invariant and which is determined variationally.
We find that the expression of $H$ in terms of this new set of
canonical operators essentially consists in a kinetic term containing
a much smaller gap, an RKKY term without any modification, and
a residual Kondo interaction which is only effective at low
energies. With these optimal operators and neglecting the spin 
correlations, we construct SC charged modes using an ansatz that 
exploits all the symmetries of the Hamiltonian. We find that
SC modes of spin $3/2$ and $1/2$ can form, respectively, in
ferro and antiferromagnetic Kondo lattices and that its formation
is much more favored in the last case. We also find that for
suffiently strong couplings, these modes must strongly renormalize
the low-energy charged spectrum and that, beyond a critical
coupling, they would condensate in the ground state producing
an instability and, consequently, a transition to a new phase.
Since the coupling of spin waves to conduction electrons
to form SC modes influences the spin correlations (which
are the main source of low-energy entropy), this
phase transition should be accompanied by important
changes in the specific heat.

\section {First transformation: interacting vacuum state}

We shall study a KLM consisting in a conduction band coupled to
a lattice of $s=1/2$ local moments by an exchange interaction.
The Hamiltonian is given by
\begin{equation}
H=\sum_{{\bf k},\alpha}\varepsilon_{{\bf k}} 
c^{\dag}_{{\bf k}\alpha} c_{{\bf k}\alpha}+
J \sum_i {\bf S}_{ei} {\bf S}_{fi}\, ,
\label{klh}
\end{equation}
where ${\bf S}_{ei}$ and ${\bf S}_{fi}$ are the spin of the
conduction electrons and the local spin at site
$i$ respectively. ${\bf S}_{ei}=1/2 \sum_{\alpha,\beta}
c^{\dag}_{i\alpha}{\bf \sigma}_{\alpha
\beta} c_{i\beta}$, where $c^{\dag}_{i\alpha}$ are electronic
operators in the site (Wannier) representation.
If ${\bf R}_{i}$ is the position of the $i$ site, these 
operators are given, for a lattice with $N$ sites, by
$c^{\dag}_{i\alpha}=N^{-1/2}\sum_{{\bf k}}{\rm e}^{-i
{\bf R}_i{\bf k}} c^{\dag}_{{\bf k}\alpha}$.

We will consider the case of a half-filled conduction band with
a symmetric density of states. Although the symmetry 
between electrons and holes will be very convenient to simplify the 
formalism, the scenario presented in this
paper is expected to hold in all situations around half filling.
To simplify the notation, we shall take energy units
in which the band width is normalized ($W=1$). The final formulas
that we will obtain
can, of course, be written in arbitrary units by suitably 
reintroducing $W$ using dimensional analysis. In (\ref{klh}), 
we have already subtracted the chemical potential in 
$\varepsilon_{{\bf k}}$ and, thus, $W=1$ and the electron-hole 
symmetry, imply that $-1/2\leq \varepsilon_{{\bf k}} \leq 1/2$.
Finally, to enable an analytic calculation of many quantities
throughout this paper, we will assume a constant
density of states $D$. $D$ normalized and constant, 
and $W=1$ imply $D=1$.

The fundamental operators $c_{{\bf k}\alpha}$, ${\bf S}_{fi}$ are
characterized by the following algebraic properties:
\begin{eqnarray}
\left \{ c_{{\bf k}\alpha},c^{\dag}_{{\bf k'}\beta} \right \}&=&
\delta_{{\bf k} {\bf k'}}\delta_{\alpha\beta}\; , \;
\left \{ c_{{\bf k}\alpha},c_{{\bf k'}\beta} \right \} =0\, ,
\label{anticomrel} \\
\left[ S^{\alpha}_{fi},S^{\beta}_{fj} \right] &=&i \delta_{ij}
\sum_{\gamma}\epsilon_{\alpha\beta\gamma}
S^{\gamma}_{fi}\, , \label{comrel} \\
{\bf S}^{\dag}_{fi}&=&{\bf S}_{fi}\, ,\label{selfad} \\
{\bf S}^2_{fi}&=&3/4,\label{ssquare} \\
\left[ c_{{\bf k}\alpha},{\bf S}_{fi} \right]
&=&0\, . \label{comcs}
\end{eqnarray}

As it is well-known, Eqs.\ (\ref{comrel}) and (\ref{selfad}) can
be satisfied by writing
${\bf S}_{fi}=1/2\sum_{\alpha,\beta} f^{\dag}_{i\alpha}{\bf \sigma}_{\alpha
\beta} f_{i\beta}$,
where $f^{\dag}_{i\alpha}$ and $f_{i\alpha}$ are creation and
annihilation operators of $f$ electrons satisfying canonical
anticommutation relations. It can be proved (and is physically
obvious) that the condition (\ref{ssquare}) is equivalent to 
demanding that each site be occupied by one and only 
one $f$ electron. 
As it will soon become clear, instead of using 
the ${\bf S}_{fi}$ operators, it is much more convenient to work
with the following equivalent set of operators:
\begin{eqnarray}
s_{1,i} & \equiv & \sqrt{2} \left( S^{x}_{fi}+i
S^{y}_{fi} \right)
= \sqrt{2} f^{\dag}_{i\uparrow}
f_{i\downarrow}\, , \nonumber \\
s_{0,i} & \equiv & 2 S^{z}_{fi}=
f^{\dag}_{i\uparrow} f_{i\uparrow}
-f^{\dag}_{i\downarrow} f_{i\downarrow}\, , \label{smalls} \\
s_{-1,i} & \equiv & \sqrt{2}
\left( S^{x}_{fi}-i S^{y}_{fi} \right)
= \sqrt{2} f^{\dag}_{i\downarrow} f_{i\uparrow}\, . \nonumber
\end{eqnarray}

It is straightforward to see that these operators,
acting on states containing a single $f$ electron per lattice
site, satisfy the following multiplication table:
\begin{equation}
\begin{array}{rlrl}
s_{0,i} s_{1,i}&=s_{1,i}&\;\;\;\;\;\;\;\;
s_{0,i}s_{-1,i}&=-s_{-1,i} \\
s_{1,i} s_{0,i}&=-s_{1,i}&      s_{-1,i}s_{0,i}&=s_{-1,i} \\
s_{1,i} s_{-1,i}&=1+s_{0,i}&   s^{2}_{1,i}&=s^{2}_{-1,i}=0 \\
s_{-1,i} s_{1,i}&=1-s_{0,i}&   s^{2}_{0,i}&=1
\end{array}
\label{mtable}
\end{equation}
as well as the properties
\begin{equation}
s^{\dag}_{1,i}=s_{-1,i}\; , \; s^{\dag}_{0,i}=
s_{0,i}\, , \label{adjoints}
\end{equation}
\begin{equation}
\left[ s_{l,i},s_{l',j} \right]=0 \;\;\; \forall i\neq j\, .
\label{comij}
\end{equation}

Eqs.\ (\ref{mtable})--(\ref{comij}) can also be regarded as
the defining conditions of the algebra of local moments 
since, as it can be readily checked, they imply Eqs.\
(\ref{comrel})--(\ref{ssquare}).

The multiplication table (\ref{mtable}) will be extremely
useful since it will enable us to reduce  products of
$s_l$ operators corresponding to the same lattice site.
For example,
\begin{equation}
s_{0,5}s_{-1,7}s_{1,5}s_{1,7}=s_{1,5}-s_{1,5}s_{0,7}\, .
\label{exam}
\end{equation}

In order to exploit the symmetry of the Hamiltonian 
(\ref{klh}) under lattice translations, it is convenient
to work with all operators in the momentum representation:
\begin{equation}
c^{\dag}_{{\bf k}\alpha}=N^{-1/2} \sum_i {\rm e}^{i {\bf R}_i
{\bf k}} c^{\dag}_{i\alpha}\;\; , \;\;
s_{l,{\bf k}}=N^{-1/2} \sum_i {\rm e}^{i {\bf R}_i
{\bf k}} s_{l,i}\, ,
\end{equation}
where ${\bf k}$ is always in the first Brillouin zone.
The $s_{l,{\bf k}}$ operators satisfy the following equations,
which will later be of much use:
\begin{eqnarray}
\left [ s_{0,{\bf k}},s_{\pm 1,
{\bf k}'} \right] &=&\pm 2 N^{-1/2}
s_{\pm 1,{\bf k}+{\bf k}'}\, , \nonumber\\
\left [ s_{1,{\bf k}},s_{-1,{\bf k}'} \right] &=&2 N^{-1/2}
s_{0,{\bf k}+{\bf k}'}\, ,\label{sscom} \\
\left [ s_{l,{\bf k}},s_{l,{\bf k}'} \right] &=&0\, ,\nonumber\\
s^{\dag}_{l,{\bf k}}&=&s_{-l,-{\bf k}}\, .
\end{eqnarray}

Another tool which we are going to use constantly is the 
development of a product of two $s_l$ operators in the
momentum representation; here is an example:
\begin{equation}
s_{-1,{\bf k}} s_{1,{\bf k'}}=\delta({\bf k}+{\bf k'})
-N^{-1/2}s_{0,{\bf k}+{\bf k'}}+N^{-1}\sum_{i\neq j}
{\rm e}^{i({\bf R}_i {\bf k}+{\bf R}_j {\bf k'})}
s_{-1,i}s_{1,j}\, .
\label{dev2}
\end{equation}

Finally, in the momentum representation, the KLH (\ref{klh}) is given by
\begin{equation}
H=\varepsilon_{{\bf k}} c^{\dag}_{{\bf k}\alpha} c_{{\bf k}\alpha}
+\frac{J}{4 N^{1/2}} \left [ (c^{\dag}_{{\bf k}\uparrow}
c_{{\bf k}'\uparrow}-c^{\dag}_{{\bf k}\downarrow}
c_{{\bf k}'\downarrow}) s_{0,{\bf k}'-{\bf k}}+ \sqrt{2}
c^{\dag}_{{\bf k}\uparrow} c_{{\bf k}'\downarrow}
s_{-1,{\bf k}'-{\bf k}}+\sqrt{2} c^{\dag}_{{\bf k}\downarrow}
c_{{\bf k}'\uparrow}s_{1,{\bf k}'-{\bf k}} \right] ,
\label{klh2}
\end{equation}
where, as in many equations throughout this paper, sum over
repeated indexes in each term of the equation 
is implicitly understood.

We want to study this Hamiltonian for small values of $J/W$,
which is the region characteristic of HF systems. In the absence
of interaction, all the spin configurations will be degenerate,
and the ground (or vacuum) state of the system $|\Phi_0\rangle$ 
will consist in a completely disordered spin lattice plus a
Fermi sea containing all electrons with negative energy.
The charged excitations of this system are electrons and holes
consisting, respectively, in adding or removing an electron
from this Fermi sea. To simplify the notation, ${\bf q}$
(${\bf p}$) will always denote a wave vector located below
(above) the Fermi surface ($\varepsilon_{\bf q}< 0$,
$\varepsilon_{\bf p}> 0$), while ${\bf k}$ will be a
wave vector without such restrictions. As it is obvious
and well-known, the creation and annihilation operators
corresponding to electrons and holes are given by
\begin{equation}
e^{\dag}_{{\bf p}\alpha}=c^{\dag}_{{\bf p}\alpha}\; , \;
e_{{\bf p}\alpha}=c_{{\bf p}\alpha}\; ,\;
h^{\dag}_{{\bf q}\alpha}=c_{{\bf -q},-\alpha}\; ,\;
h_{{\bf q}\alpha}=c^{\dag}_{{\bf -q},-\alpha}\, ,
\label{eandh}
\end{equation}
and the non-interacting vacuum state can be written as
\begin{equation}
|\Phi_0\rangle =\sum_{\alpha_{i}=\uparrow,\downarrow} 
C^{(0)}_{\alpha_{1},\ldots,\alpha_{N}} f^{\dag}_{1\alpha_{1}} 
\ldots f^{\dag}_{N\alpha_{N}}|\Psi_0\rangle ,
\end{equation}
where $|\Psi_0\rangle$ is a state annihilated by 
$e_{{\bf p}\alpha}$ and $h_{{\bf q}\alpha}$, and the coefficients
$C^{(0)}_{\alpha_{1},\ldots,\alpha_{N}}$ are given by
$C^{(0)}_{\alpha_{1},\ldots,\alpha_{N}}=2^{-N/2}
{\rm e}^{i R_{\alpha_{1} \ldots \alpha_{N}}}$,
where $R_{\alpha_{1} \ldots \alpha_{N}}$ are random phases,
as it corresponds to a completely disordered spin lattice.

The KLH (\ref{klh2}) is given, in terms of
the operators (\ref{eandh}), by
\begin{eqnarray}
&&H=( {\textstyle \sum_q} 2 \varepsilon_{\bf q})+
\varepsilon_{\bf p}e^{\dag}_{{\bf p}\alpha}e_{{\bf p}\alpha}
-\varepsilon_{\bf q}h^{\dag}_{{\bf q}\alpha}h_{{\bf q}\alpha}\\
&&+\frac{J}{4N^{1/2}}\left[ (e^{\dag}_{{\bf p}\uparrow}
e_{{\bf p'}\uparrow}-e^{\dag}_{{\bf p}\downarrow}
e_{{\bf p'}\downarrow})s_{0,{\bf p'}-{\bf p}}
+\sqrt{2}e^{\dag}_{{\bf p}\uparrow}e_{{\bf p'}\downarrow}
s_{-1,{\bf p'}-{\bf p}}+\sqrt{2}e^{\dag}_{{\bf p}\downarrow}
e_{{\bf p'}\uparrow}s_{1,{\bf p'}-{\bf p}} \right] \label{epest}\\
&&+\frac{J}{4N^{1/2}}\left[ (h^{\dag}_{{\bf q}\uparrow}
h_{{\bf q'}\uparrow}-h^{\dag}_{{\bf q}\downarrow}
h_{{\bf q'}\downarrow})s_{0,{\bf q'}-{\bf q}}
-\sqrt{2}h^{\dag}_{{\bf q}\uparrow}h_{{\bf q'}\downarrow}
s_{-1,{\bf q'}-{\bf q}}-\sqrt{2}h^{\dag}_{{\bf q}\downarrow}
h_{{\bf q'}\uparrow}s_{1,{\bf q'}-{\bf q}}\right] \label{hphst}\\
&&+\frac{J}{4N^{1/2}}\left[
(e^{\dag}_{{\bf p}\uparrow}h^{\dag}_{{\bf q}\downarrow}-
e^{\dag}_{{\bf p}\downarrow}h^{\dag}_{{\bf q}\uparrow})
s_{0,{\bf -q}-{\bf p}}+
\sqrt{2}e^{\dag}_{{\bf p}\uparrow}h^{\dag}_{{\bf q}\uparrow}
s_{-1,{\bf -q}-{\bf p}}+
\sqrt{2}e^{\dag}_{{\bf p}\downarrow}h^{\dag}_{{\bf q}\downarrow}
s_{1,{\bf -q}-{\bf p}}\right]\label{ppt}\\
&&+\frac{J}{4N^{1/2}} \left[ (h_{{\bf q}\downarrow}e_{{\bf p}\uparrow}-
h_{{\bf q}\uparrow}e_{{\bf p}\downarrow})s_{0,{\bf q}+{\bf p}}+
\sqrt{2}h_{{\bf q}\uparrow}e_{{\bf p}\uparrow}
s_{1,{\bf q}+{\bf p}}+
\sqrt{2}h_{{\bf q}\downarrow}e_{{\bf p}\downarrow}
s_{-1,{\bf q}+{\bf p}} \right]\, . \label{mmt}
\end{eqnarray}

This expression makes clear that $|\Phi_0\rangle$ is not an 
eigenstate (and, hence, not the ground state) of $H$
when $J\neq 0$, because the terms (\ref{ppt}) do not
annihilate $|\Phi_0\rangle$ but create additional 
electron-hole excitations accompanied by a spin fluctuation.

To determine the elementary excitations of the interacting
system, we must first find the interacting vacuum state 
$|\Phi\rangle$. Our approach will consist
in determining a unitary transformation that maps
the basic operators $c_{{\bf k}\alpha}$, $s_{l,{\bf k}}$
into a new set $\tilde{c}_{{\bf k}\alpha}$, 
$\tilde{s}_{l,{\bf k}}$ such that $|\Phi\rangle$
can be written as
\begin{equation}
|\Phi\rangle =\sum_{\alpha_{i}=\uparrow,\downarrow}
C_{\alpha_{1},\ldots,\alpha_{N}} \tilde{f}^{\dag}_{1\alpha_{1}}
\ldots \tilde{f}^{\dag}_{N\alpha_{N}}|\Psi\rangle ,
\label{igs}
\end{equation}
where $\tilde{e}_{{\bf p}\alpha}|\Psi\rangle =
\tilde{h}_{{\bf q}\alpha}|\Psi\rangle =0$, and the new local
Fermi operators $\tilde{f}_{i,\alpha}$ are related to
$\tilde{s}_{l,i}$ by Eq.\ ($\tilde{\ref{smalls}}$),
(we denote by ($\tilde{\rm x}$) a former equation (x) with all
operators tilded).

The transformation is chosen to be unitary to ensure that
the new operators continue satisfying the basic,
defining algebra ($\tilde{\ref{anticomrel}}$)--($\tilde{\ref{comcs}}$).
The relationship between
the initial and transformed operators is, thus,
\begin{equation}
c^{\dag}_{{\bf k}\alpha}={\rm e}^{\tilde{T}} 
\tilde{c}^{\dag}_{{\bf k}\alpha}
{\rm e}^{-\tilde{T}}\;,\; s_{l,{\bf k}}=
{\rm e}^{\tilde{T}} \tilde{s}_{l,{\bf k}}
{\rm e}^{-\tilde{T}}\;, \;\;\;\tilde{T}^{\dag}=-\tilde{T}\,.
\label{trans1}
\end{equation}

Since we are interested, at this point, in determining a
ground state that does not break any symmetry,
we will demand that $\tilde{T}$ preserve the charge, lattice translational,
and spin rotational symmetries. The simplest operator 
involving fermions that satisfies this condition is
\begin{equation}
\tilde{T}=\frac{J}{4N^{1/2}}\tilde{T}({\bf k},{\bf k'})\left [ 
(\tilde{c}^{\dag}_{{\bf k}\uparrow}
\tilde{c}_{{\bf k'}\uparrow}-\tilde{c}^{\dag}_{{\bf k}\downarrow}
\tilde{c}_{{\bf k'}\downarrow})\tilde{s}_{0,{\bf k'}-{\bf k}}
+\sqrt{2}\tilde{c}^{\dag}_{{\bf k}\uparrow}
\tilde{c}_{{\bf k'}\downarrow}\tilde{s}_{-1,{\bf k'}-{\bf k}}+
\sqrt{2}\tilde{c}^{\dag}_{{\bf k}\downarrow}
\tilde{c}_{{\bf k'}\uparrow}\tilde{s}_{1,{\bf k'}-{\bf k}}\right] ,
\label{ttrans}
\end{equation}
where 
\begin{equation}
\tilde{T}^{\star}({\bf k'},{\bf k})=-\tilde{T}({\bf k},{\bf k'})
\label{starm}
\end{equation}
in order to satisfy $\tilde{T}^{\dag}=-\tilde{T}$. 
Apart from this condition,
the function $\tilde{T}({\bf k},{\bf k'})$ is arbitrary, and it
will be determined by demanding that the state (\ref{igs})
be the actual interacting vacuum state.

For small values of $J$, the new operators are expected to be
slight deformations of the initial ones. Thus, we will expand
Eq.\ (\ref{trans1}) in powers of $\tilde{T}$, retaining the first order
terms only:
\begin{equation}
c^{\dag}_{{\bf k}\alpha}=\tilde{c}^{\dag}_{{\bf k}\alpha}+
[ \tilde{T},\tilde{c}^{\dag}_{{\bf k}\alpha}]\; ,\;
s_{l,{\bf k}}=\tilde{s}_{l,{\bf k}}+
[ \tilde{T},\tilde{s}_{l,{\bf k}}]\,.
\end{equation}

These equations are the basis of the forecoming calculations.
Using Eqs.\ ($\tilde{\ref{anticomrel}}$) and 
($\tilde{\ref{sscom}}$),
we can write them in a more explicit form:
\begin{eqnarray}
c^{\dag}_{{\bf k}\uparrow(\downarrow)}&=&\tilde{\alpha} ({\bf k})
\left\{ \tilde{c}^{\dag}_{{\bf k}\uparrow(\downarrow)}
+\frac{J}{4N^{1/2}}\tilde{T}({\bf k'},{\bf k})\left[ +(-)
\tilde{c}^{\dag}_{{\bf k'}\uparrow(\downarrow)}
\tilde{s}_{0,{\bf k}-{\bf k'}}+\sqrt{2}
\tilde{c}^{\dag}_{{\bf k'}\downarrow(\uparrow)}
\tilde{s}_{1(-1),{\bf k}-{\bf k'}}\right]\right\},
\label{cpno}\\
c_{{\bf k}\uparrow(\downarrow)}&=&\tilde{\alpha} ({\bf k})
\left\{ \tilde{c}_{{\bf k}\uparrow(\downarrow)}
+\frac{J}{4N^{1/2}}\tilde{T}^{\star}({\bf k'},{\bf k})\left[ +(-)
\tilde{c}_{{\bf k'}\uparrow(\downarrow)}
\tilde{s}_{0,{\bf k'}-{\bf k}}+\sqrt{2}
\tilde{c}_{{\bf k'}\downarrow(\uparrow)}
\tilde{s}_{-1(1),{\bf k'}-{\bf k}}\right]\right\},
\label{cno}\\
s_{0,{\bf k}}&=&\tilde{s}_{0,{\bf k}}+\frac{J}{2N}
\tilde{T}({\bf k_1},{\bf k_2})\left[ \sqrt{2}
\tilde{c}^{\dag}_{{\bf k_1}\uparrow}\tilde{c}_{{\bf k_2}\downarrow}
\tilde{s}_{-1,{\bf k_2}-{\bf k_1}+{\bf k}}-\sqrt{2}
\tilde{c}^{\dag}_{{\bf k_1}\downarrow}\tilde{c}_{{\bf k_2}\uparrow}
\tilde{s}_{1,{\bf k_2}-{\bf k_1}+{\bf k}}\right],\\
s_{1(-1),{\bf k}}&=&\tilde{s}_{1(-1),{\bf k}}+(-)\frac{J}{2N}
\tilde{T}({\bf k_1},{\bf k_2})\left[ (
\tilde{c}^{\dag}_{{\bf k_1}\uparrow}\tilde{c}_{{\bf k_2}\uparrow}-
\tilde{c}^{\dag}_{{\bf k_1}\downarrow}\tilde{c}_{{\bf k_2}\downarrow})
\tilde{s}_{1(-1),{\bf k_2}-{\bf k_1}+{\bf k}}\right.\nonumber \\
&&\;\;\;\;\;\;\;\;\;\;\;\;\;\;\;\;\;\;\;\;\;\;\;\;\;\;\;\;\;\;\;\;\;
\;\;\;\;\;\;\;\;\;\;\;\;\;\;\;\;\;\;\;\;\;\;\;\;\;\;\;\left.
-\sqrt{2}\tilde{c}^{\dag}_{{\bf k_1}\uparrow(\downarrow)}
\tilde{c}_{{\bf k_2}\downarrow(\uparrow)}
\tilde{s}_{0,{\bf k_2}-{\bf k_1}+{\bf k}}\right],
\label{s1no}
\end{eqnarray}
where the coefficient $\tilde{\alpha}({\bf k})$ in Eqs.\ (\ref{cpno})
and (\ref{cno}) is, for the moment, equal to one.

With these equations, we can readily express the KLH
(\ref{klh2}) in terms of the transformed operators
$\tilde{e}_{{\bf p}\alpha}$, $\tilde{h}_{{\bf q}\alpha}$, and
$\tilde{s}_{l,i}$. To carry out this program, we have to:
\renewcommand{\theenumi}{\roman{enumi}}
\begin{enumerate}
\item Substitute Eqs.\ (\ref{cpno})--(\ref{s1no}) in the expression
of the Hamiltonian (\ref{klh2}),
\item write the $\tilde{c}_{{\bf k}\alpha}$ operators in terms
of $\tilde{e}_{{\bf p}\alpha}$ and $\tilde{h}_{{\bf q}\alpha}$
using Eq.\ ($\tilde{\ref{eandh}}$),
\item normal-order these fermionic operators using the
canonical anticommutation relations ($\tilde{\ref{anticomrel}}$),
\item and contract the $\tilde{s}_{l,i}$ operators corresponding
to the same lattice site using the multiplication table 
($\tilde{\ref{mtable}}$) as in Eqs.\ ($\tilde{\ref{exam}}$) 
and ($\tilde{\ref{dev2}}$).
\end{enumerate}

The execution of this program is straightforward but somewhat
lengthy, since many terms are generated. To easily identify
the procedence of each term in the following discussion, we will 
symbolically represent Eqs.\ (\ref{cpno})--(\ref{s1no}) by
$c^{\dag}=\tilde{c}^{\dag}+\delta\tilde{c}^{\dag}$,
$c=\tilde{c}+\delta\tilde{c}$,
and $s=\tilde{s}+\delta\tilde{s}$.

The kinetic and Kondo parts of $H$ have the structures $c^{\dag}c$
and $c^{\dag}cs$ respectively.
We will say, for example, that a particular term comes from
$(\tilde{c}^{\dag}|\delta\tilde{c}|\tilde{s})$, meaning that it
has been generated by executing the steps (ii)--(iv) on
the part of the development of $c^{\dag}cs$ with that structure.

We will now determine $\tilde{T}({\bf k},{\bf k'})$. 
Writing $\tilde{T}$
in Eq.\ (\ref{ttrans}) with the electron and hole 
operators ($\tilde{\ref{eandh}}$), we get
that $\tilde{T}=\tilde{T}_1+\tilde{T}_2$, where $\tilde{T}_1$ 
and $\tilde{T}_2$ have the structures
\begin{eqnarray}
\tilde{T}_1&=&\frac{J}{4N^{1/2}}\left[ \tilde{T}({\bf p},{\bf q})
\tilde{e}^{\dag}_{\bf p}\tilde{h}^{\dag}_{\bf -q}
\tilde{s}_{{\bf q}-{\bf p}}+\tilde{T}({\bf q},{\bf p})
\tilde{h}_{\bf -q}\tilde{e}_{\bf p}
\tilde{s}_{{\bf p}-{\bf q}}\right] , \label{1t}\\
\tilde{T}_2&=&\frac{J}{4N^{1/2}}\left[ \tilde{T}({\bf p},{\bf p'})
\tilde{e}^{\dag}_{\bf p}\tilde{e}_{\bf p'}
\tilde{s}_{{\bf p'}-{\bf p}}
-\tilde{T}({\bf q},{\bf q'})\tilde{h}^{\dag}_{\bf -q'}
\tilde{h}_{\bf -q}\tilde{s}_{{\bf q'}-{\bf q}}\right] .
\label{2t}
\end{eqnarray}

The unitary transformation ${\rm e}^{\tilde{T}}$ 
can be written, to leading order in $J$, as the product 
of ${\rm e}^{\tilde{T}_2}$ and ${\rm e}^{\tilde{T}_1}$.
Since $\tilde{T}_2|\Phi\rangle =0$, ${\rm e}^{\tilde{T}_2}$ 
does not contribute to
change the vacuum state $|\Phi\rangle$, and it can be taken to be
the identity without loss of generality:
\begin{equation}
\tilde{T}({\bf p},{\bf p'})=
\tilde{T}({\bf q},{\bf q'})=0\;\;\;\;\forall
{\bf p},{\bf p'},{\bf q},{\bf q'}\, .
\end{equation}

The presence of terms of the form $\tilde{e}^{\dag}\tilde{h}^{\dag}
\tilde{s}$ in $H$ would imply that $|\Phi\rangle$ is not an
eigenstate and, hence, not the ground state of the system. 
In fact, by demanding the cancellation of these terms, we will 
be able to determine $\tilde{T}({\bf p},{\bf q})$. To leading order 
in $J$, this condition is given by
\begin{equation}
(\tilde{\ref{ppt}})-\tilde{T}({\bf p},{\bf -q})(\varepsilon_{\bf p}-
\varepsilon_{\bf q})\times(\tilde{\ref{ppt}})=0\, ,
\label{fot}
\end{equation}
where the first term comes from $(\tilde{c}^{\dag}|\tilde{c}|
\tilde{s})$ and the second from $(\tilde{c}^{\dag}|\delta\tilde{c})$
and $(\delta\tilde{c}^{\dag}|\tilde{c})$. This condition and
Eq.\ (\ref{starm}) imply that 
\begin{equation}
\tilde{T}({\bf p},{\bf q})=-\tilde{T}({\bf q},{\bf p})=
\frac{1}{\varepsilon_{\bf p}-\varepsilon_{\bf q}}\, .
\label{tpq0}
\end{equation}

Since $\varepsilon_{\bf p}>0$ and $\varepsilon_{\bf q}<0$, the
denominator in this equation is always positive. However, when
${\bf p}$ and ${\bf q}$ are both very close to the Fermi surface
(low-energy region),
this denominator tends to vanish and $\tilde{T}({\bf p},{\bf q})$ 
becomes very large. This means that, in this region, higher order
terms in $J$ can give important contributions to
$\tilde{e}^{\dag}\tilde{h}^{\dag}\tilde{s}$. A careful inspection 
of these terms leads to the conclusion that their dominant
contribution at low energies comes from $(\delta\tilde{c}^{\dag}|
\delta\tilde{c}|\tilde{s})$ and is given by
\begin{equation}
-\frac{3J^2}{16}\left[\frac{1}{N}\sum_{\bf p'}
\tilde{T}({\bf p'},{\bf q})+\frac{1}{N}\sum_{\bf q'}
\tilde{T}({\bf p},{\bf q'})\right]\tilde{T}({\bf p},
{\bf q})\times(\tilde{\ref{ppt}}) \, . \label{hot}
\end{equation}

Assuming a constant density of states ($D=1$), the sums in 
this equation can be straightforwardly calculated:
\begin{equation}
\frac{1}{N}\sum_{\bf p'}\tilde{T}({\bf p'},{\bf q})=\int_0^{1/2}
\frac{d\varepsilon_{\bf p'}}{\varepsilon_{\bf p'}-
\varepsilon_{\bf q}}\sim\ln\frac{-1}{2\varepsilon_{\bf q}}\;\;,\;\;
(-\varepsilon_{\bf q}\ll 1/2)\, .
\end{equation}

Thus, the terms (\ref{hot}) diverge as $\varepsilon_{\bf p}$ and
$\varepsilon_{\bf q}$ vanish and, consequently, the cancellation
of the contributions to $\tilde{e}^{\dag}\tilde{h}^{\dag}\tilde{s}$
is not achieved.

This problematic state-of-affairs can be overcome by introducing
a low-energy regulator in Eq.\ (\ref{tpq0}), namely,
\begin{equation}
\tilde{T}({\bf p},{\bf q})=-\tilde{T}({\bf q},{\bf p})=
\frac{1}{\varepsilon_{\bf p}-\varepsilon_{\bf q}+\tilde{\eta}}\, .
\label{tpq}
\end{equation}
where $\tilde{\eta}$ is a small positive 
function. Since at high energies
this function is neglible compared to $\varepsilon_{\bf p}$ and
$-\varepsilon_{\bf q}$, it will only play a role at very low
energies and, therefore, we can approximately take it as a constant 
equal to its zero-energy value.
Obviously, $\tilde{\eta}$ will be determined
by demanding that the sum of all contributions to the terms 
$\tilde{e}^{\dag}\tilde{h}^{\dag}\tilde{s}$ [(\ref{fot}) and
(\ref{hot})] vanish when $\varepsilon_{\bf p}=
\varepsilon_{\bf q}=0$, namely, 
\begin{equation}
\left[ 1-\frac{3J^2}{8\tilde{\eta}}\ln\frac{1}{2\tilde{\eta}}\right]
\times(\tilde{\ref{ppt}})
\left|_{\varepsilon_{\bf p}=\varepsilon_{\bf q}=0}\right.=0\, .
\end{equation}
Thus, $\tilde{\eta}$ is given, as a function of $J$, by
\begin{equation}
\tilde{\eta}=\frac{3J^2}{8}\ln\frac{1}{2\tilde{\eta}}\,.
\label{regulator}
\end{equation}

Substituting (\ref{tpq}),
we see that the terms (\ref{hot}) are neglegible compared
to the first order terms (\ref{fot}) in the high-energy region.
Since in this region (\ref{tpq}) essentially coincides with
(\ref{tpq0}), Eq.\ (\ref{fot}) will continue to hold to leading order
in $J$ at high energies. In summary, with $\tilde{T}({\bf p},{\bf q})$ 
and $\tilde{\eta}$ given, respectively, by (\ref{tpq}) and
(\ref{regulator}), we accomplish our objective of cancelling the 
terms $\tilde{e}^{\dag}\tilde{h}^{\dag}\tilde{s}$ for all energies
and, since $H$ is a hermitian operator, this also implies the
cancellation of the terms $\tilde{h}\tilde{e}\tilde{s}$.

To verify that unitarity is indeed preserved by this transformation,
we can calculate the canonical properties 
(\ref{anticomrel}) and (\ref{mtable})
in terms of the transformed operators [using Eqs.\ 
(\ref{cpno})--(\ref{s1no}),
($\tilde{\ref{anticomrel}}$) and ($\tilde{\ref{mtable}}$)],
and check whether they continue to be satisfied. This calculation
yields, for the terms proportional to 1,
\begin{eqnarray}
\left \{ c_{{\bf k}\alpha},c^{\dag}_{{\bf k'}\beta} \right \}&=&
[\tilde{\alpha}({\bf k})]^{-2}\delta_{{\bf k},{\bf k'}}
\delta_{\alpha\beta}\, , \\
s_{0,{\bf k}}s_{0,{\bf k'}}&=&\tilde{\sigma}^{-2}
\delta_{{\bf k},{\bf -k'}}\, , 
\end{eqnarray}
where
\begin{eqnarray}
\tilde{\alpha}({\bf k})&=&\left[ 1+\frac{3J^2}{16N}\sum_{\bf k'}
\tilde{T}^2({\bf k'},{\bf k})\right]^{-1/2}=
\left[1+\frac{3J^2}{16}\left(\frac{1}{|\varepsilon_{\bf k}|+
\tilde{\eta}}-\frac{1}{|\varepsilon_{\bf k}|+1/2}\right)
\right]^{-1/2} , \label{alpha} \\
\tilde{\sigma}&=&\left[ 1+\frac{J^2}{N^2}\sum_{{\bf p},{\bf q}}
\tilde{T}^2({\bf p},{\bf q})\right]^{-1/2}=\left[1+J^2
\ln\frac{1}{4\tilde{\eta}}\right]^{-1/2} .
\end{eqnarray}
Since both $\tilde{\alpha}({\bf k})$ and $\tilde{\sigma}$ tend 
to 1 for small values of $J$, the canonical algebraic properties
are essentially preserved. This would certainly not be the
case if $\tilde{\eta}=0$ and, thus, the introduction of a
regulator is not only necessary to cancel the
$\tilde{e}^{\dag}\tilde{h}^{\dag}\tilde{s}$ terms
at low energies, but also to achieve unitarity. The departure
of $\tilde{\sigma}$ from 1 is very small and can be neglected.
In the case of $\tilde{\alpha}({\bf k})$, however, this departure
is not that small at low energies. If ${\bf k}_F$ is a wave
vector of the Fermi surface, $\tilde{\alpha}({\bf k}_F)\sim
1-1/(4\ln\frac{1}{2\tilde{\eta}})$. For small $J$, the
second term of this equation is certainly neglegible
compared to 1, but it is large compared to $J$. Thus, to 
satisfy more accurately the canonical
anticommutation relations for the two sets of operators,
we will substitute the factor $\tilde{\alpha}({\bf k})$ in Eqs.\
(\ref{cpno}) and (\ref{cno}) (which up till now was set 
equal to one) by the function (\ref{alpha}).

Having determined completely the relationship between the
initial and transformed operators [Eqs.\ 
(\ref{cpno})--(\ref{s1no}), (\ref{tpq}), (\ref{regulator}),
and (\ref{alpha})], we can now proceed to calculate $H$
in terms of $\tilde{e}_{{\bf p}\alpha}$, 
$\tilde{h}_{{\bf q}\alpha}$, and $\tilde{s}_{l,i}$ following
the steps (i)--(iv) detailed after Eq.\ (\ref{s1no}). 
From the many contributions to each particular term, we shall
only retain the dominant ones for small $J$. Next, we summarize
the results obtained, indicating where the dominant contributions
stem from:

\begin{itemize}
\item $\tilde{H}_{\em kinetic}$: The main contributions
to the terms $\tilde{e}^{\dag}\tilde{e}$ and $\tilde{h}^{\dag}
\tilde{h}$ come from $(\tilde{c}^{\dag}|\tilde{c})$,
$(\delta\tilde{c}^{\dag}|\delta\tilde{c})$,
$(\tilde{c}^{\dag}|\delta\tilde{c}|\tilde{s})$,
and $(\delta\tilde{c}^{\dag}|\tilde{c}|\tilde{s})$, and
are given by
\begin{equation}
\tilde{H}_{\em kinetic}=\sum_{{\bf p},\alpha}
\tilde{E}(\varepsilon_{\bf p})\tilde{e}^{\dag}_{{\bf p}\alpha}
\tilde{e}_{{\bf p}\alpha}+\sum_{{\bf q},\alpha}
\tilde{E}(-\varepsilon_{\bf q})\tilde{h}^{\dag}_{{\bf q}\alpha}
\tilde{h}_{{\bf q}\alpha}\, ,
\end{equation}
where
\begin{equation}
\tilde{E}(\varepsilon_{\bf p})=\varepsilon_{\bf p}-\frac{3J^2}{16N}
\sum_{\bf q}\left[\left(\varepsilon_{\bf p}\tilde{\alpha}^2({\bf p})-
\varepsilon_{\bf q}\tilde{\alpha}^2({\bf q})\right)
\tilde{T}({\bf p},{\bf q})-2\tilde{\alpha}({\bf p})
\tilde{\alpha}({\bf q})\right]\tilde{T}({\bf p},{\bf q})\,.
\end{equation}
Assuming a constant density of states ($D=1$), this equation can 
be developed into
\begin{equation}
\tilde{E}(\varepsilon)=\varepsilon+\frac{3J^2}{16}\left[\left(1-
\frac{3J^2}{16}\frac{1}{\varepsilon+\tilde{\eta}}\right)\ln
\frac{\varepsilon+1/2}{\varepsilon+\tilde{\eta}}+\frac{\tilde{\eta}}
{\varepsilon+\tilde{\eta}}\right]\; , \; (\varepsilon>0)\, .
\label{disprel0}
\end{equation}
\item $\tilde{C}$: The essential 
contribution to the terms proportional to 1 comes from
normal-ordering the operator $-\sum_{{\bf q},\alpha}
\tilde{E}(-\varepsilon_{\bf q})\tilde{h}_{{\bf q}\alpha}
\tilde{h}^{\dag}_{{\bf q}\alpha}$, which yields
\begin{equation}
\tilde{C}=-\sum_{\bf q}2\tilde{E}(-\varepsilon_{\bf q})=
E_{{\Phi}_0} -\frac{3J^2}{8}N\ln 2\, ,
\end{equation}
where $E_{{\Phi}_0}=\sum_{\bf q}2\varepsilon_{\bf q}=-N/4$ is the
energy of the non-interacting vacuum state $|\Phi_0\rangle$.
\item $\tilde{H}_{\em Kondo}$: The terms
$\tilde{e}^{\dag}\tilde{h}^{\dag}\tilde{s}$ and its hermitian
conjugates $\tilde{h}\tilde{e}\tilde{s}$ are obviously absent
from $H$, since this was the condition used to determine 
$\tilde{T}({\bf p},{\bf q})$. For the terms $\tilde{e}^{\dag}\tilde{e}
\tilde{s}$ and $\tilde{h}^{\dag}\tilde{h}\tilde{s}$, the
main contributions come from $(\tilde{c}^{\dag}|\tilde{c}|
\tilde{s})$ and are given by
\begin{equation}
\tilde{H}_{\em Kondo}=
\tilde{\alpha}({\bf p})\tilde{\alpha}({\bf p'})\times 
(\tilde{\ref{epest}})+\tilde{\alpha}({\bf q})\tilde{\alpha}
({\bf q'})\times (\tilde{\ref{hphst}})\,.
\end{equation}
\item $\tilde{H}_{\em RKKY}$: The dominant contributions
to the terms $\tilde{s}_{l,i}\tilde{s}_{l',j}$ ($i\neq j$) 
are of order $J^2$ and come from 
$(\delta\tilde{c}^{\dag}|\delta\tilde{c})$,
$(\tilde{c}^{\dag}|\delta\tilde{c}|\tilde{s})$, and
$(\delta\tilde{c}^{\dag}|\tilde{c}|\tilde{s})$. This 
RKKY interaction is given by
\begin{equation}
\tilde{H}_{\em RKKY}=\frac{1}{2}\sum_{i\neq j}J_{\em RKKY}
({\bf R}_i -{\bf R}_j)\tilde{\bf S}_i\tilde{\bf S}_j\, ,
\label{hrkky}
\end{equation}
where $\tilde{\bf S}_i$ are the vector operators related to
$\tilde{s}_{l,i}$ by Eq.\ ($\tilde{\ref{smalls}}$), and
$J_{\em RKKY}({\bf R})$ is an inter-site exchange coupling
given by
\begin{equation}
J_{\em RKKY}({\bf R})=\frac{J^2}{N^2}\sum_{{\bf p},{\bf q}}
\left[\varepsilon_{\bf p}\tilde{\alpha}^2({\bf p})
\tilde{T}({\bf p},{\bf q})-2\tilde{\alpha}({\bf q})
\tilde{\alpha}({\bf p})\right]\tilde{T}({\bf p},{\bf q})
\cos[({\bf q}-{\bf p}){\bf R}]\, .
\end{equation}
or, with the approximation $\tilde{\alpha}({\bf k})\sim 1$, 
\begin{equation}
J_{\em RKKY}({\bf R})=\frac{J^2}{\Gamma^2}\int d{\bf p}\,d{\bf q}
\left(\frac{\varepsilon_{\bf p}}{\varepsilon_{\bf p}-
\varepsilon_{\bf q}+\tilde{\eta}}-2\right)\frac{\cos[({\bf q}-{\bf p})
{\bf R}]}{\varepsilon_{\bf p}-\varepsilon_{\bf q}+\tilde{\eta}}\, ,
\label{jrkky}
\end{equation}
where $\Gamma$ is the volume of the first Brillouin zone. It 
is clear from this equation that, in order to determine
$J_{\em RKKY}({\bf R})$, it is not enough to know the density
of states, but we have to specify the entire function
$\varepsilon_{\bf k}$. In one dimension, however, $W=D=1$ and
$\varepsilon_k=\varepsilon_{-k}$ imply that
$\varepsilon_k=a|k|/\pi-1/2$ where $a$ is 
the real lattice constant. This dispersion relation leads to
\begin{eqnarray}
J_{\em RKKY}(R)=\frac{J^2}{2}\int_0^{1/2}\!dx&&\int_0^{1/2}\!dy
\left(\frac{y}{1/2+y-x+\tilde{\eta}}-2\right)\times\nonumber\\
&&\times\frac{\cos[(1/2+y-x)
\pi R/a]+\cos[(1/2+y+x)\pi R/a]}{1/2+y-x+\tilde{\eta}}\, ,
\label{1drkky}
\end{eqnarray}
where, since the integrals converge as $\tilde{\eta}\to 0$, 
we can make the approximation of taking $\tilde{\eta}=0$.
From Eq.\ (\ref{jrkky}), it is clear
that $J_{\em RKKY}({\bf R})$ is a long-range, oscillating coupling.
Carrying out the integrals (\ref{1drkky}) we find that, for the 
one-dimensional case, $J_{\em RKKY}(R=\pm a)=0.26J^2$,
which shows that the exchange coupling between nearest neighbours is
antiferromagnetic.
\item $(\delta\tilde{c}^{\dag}|\delta\tilde{c})$,
$(\tilde{c}^{\dag}|\delta\tilde{c}|\tilde{s})$, and
$(\delta\tilde{c}^{\dag}|\tilde{c}|\tilde{s})$ also produce terms
of the form $\tilde{c}^{\dag}\tilde{c}\tilde{s}_i\tilde{s}_j$
($i\neq j$). The $\tilde{e}^{\dag}\tilde{h}^{\dag}
\tilde{s}_i\tilde{s}_j$ parts
and its hermitian conjugates should be 
cancelled by the $J^2$ terms of $\tilde{T}$ in the same way as
the leading orders of this transformation cancel the
$\tilde{e}^{\dag}\tilde{h}^{\dag}\tilde{s}$ 
and $\tilde{h}\tilde{e}\tilde{s}$ parts. The remaining terms, 
$\tilde{e}^{\dag}\tilde{e}\tilde{s}_i\tilde{s}_j$
and $\tilde{h}^{\dag}\tilde{h}\tilde{s}_i\tilde{s}_j$, are associated
with the interaction between charge carriers and local moments, and
it may seem at first sight that they can be neglected because
this interaction is dominated by $\tilde{H}_{\em Kondo}$. Nevertheless, 
in the presence of spin correlations without magnetic order (spin liquid),
these terms would give leading contributions to the low-energy
electronic dispersion relation because the main contribution of the
Kondo interaction vanishes 
($\langle\Phi|\tilde{s}_{l,i}|\Phi\rangle=0$). However, 
since the effect of the spin correlations
on the charged excitations will not be considered in this article, we
will not include these terms in $H$.
\end{itemize}

By carefully inspecting the rest of the terms generated by this
change of coordinates, it can be 
shown that their effect is negligible
compared to the interactions listed above. Thus,
the expression of the KLH in terms of the
new canonical operators $\tilde{e}$, $\tilde{h}$, and $\tilde{s}$
is essentially given by
\begin{equation}
H=\tilde{C}+\tilde{H}_{\em kinetic}+\tilde{H}_{\em Kondo}+
\tilde{H}_{\em RKKY}\, .
\label{1stth}
\end{equation}

\section{Spin dynamics, magnetic order,
and elementary excitations}

On the basis of Eq.\ (\ref{1stth}) we can now address the 
problem of magnetic order. To completely specify the vacuum state
$|\Phi\rangle$, we still have to determine the coefficients
$C_{\alpha_{1},\ldots,\alpha_{N}}$ of Eq.\ (\ref{igs}) which
describe the correlations among the local $\tilde{f}$ spins.
Since all the terms of $H$ containing tilded fermionic operators 
annihilate $|\Phi\rangle$, these correlations arise exclusively
as a consequence of the induced RKKY interaction. Thus, to
determine the coefficients $C_{\alpha_{1},\ldots,\alpha_{N}}$ 
we must demand the minimization of 
\begin{equation}
\langle\Phi|\tilde{H}_{\em RKKY}|\Phi\rangle =\frac{1}{2}
\sum_{i\neq j}J_{\em RKKY}({\bf R}_i-{\bf R}_j)
\langle\Phi|\tilde{\bf S}_i\tilde{\bf S}_j|\Phi\rangle .
\end{equation}
This quantity [which would be zero if $C_{\alpha_{1},\ldots,
\alpha_{N}}$ were random coefficients]
is negative, and it represents the energy decrease
due to the formation of spin correlations. The total 
ground state energy is given by
\begin{equation}
E_{\Phi}=E_{\Phi_0}-\frac{3J^2}{8}N\ln 2+
\langle\Phi|\tilde{H}_{\em RKKY}|\Phi\rangle \, ,
\end{equation}
where $E_{\Phi_0}$ is the energy of the non-interacting Fermi
sea, and the second and third terms represent the 
energy decreases due to the direct Kondo interaction and the 
induced spin correlations respectively.

Thus, within the approach presented in this paper, the 
determination of the magnetic structure of the vacuum state
for small couplings is reduced to 
finding the ground state of a Heisenberg 
Hamiltonian with an RKKY coupling given by Eq.\ (\ref{jrkky}).
This is, of course, a very hard task which we do not intend
to undertake in this article. However, it should be noted that
the situation corresponding to $\tilde{H}_{\em RKKY}$ is
fundamentally different from that of a typical Heisenberg 
Hamiltonian coupling only nearest neighbours, because a
long-range, oscillating coupling like (\ref{jrkky}) will
inevitably lead to strong frustration and spin fluctuations,
which tend to inhibit magnetic order and to favor the formation of 
a singlet RVB-like spin liquid,\cite{anderson,anderson2} or an
antiferromagnet with strongly quenched local moments.

The magnetic nature of
the ground state has been usually attributed to the competition
between the Kondo and RKKY interactions, establishing a close
parallelism with the discussion of the 
two-impurity case:\cite{jones} For weak couplings, the RKKY 
interaction (which is proportional to $J^2$) dominates and the 
two impurities order magnetically, while for stronger couplings,
the formation of a Kondo singlet at each impurity site
[with binding energy proportional to $\exp(-1/DJ)$] is 
energetically favored and magnetic order is destroyed. Between 
these two extreme cases, we would have intermediate situations 
in which the local moments are only partially compensated.

As we mentioned in the introduction, 
the extension of this approach from
the two-impurity case to the Kondo lattice is very
unclear due to the Nozi\`eres exhaustion problem.
Another objection that can be raised is that this 
mechanism of forming Kondo singlets to compensate the local
moments would not be valid for $J<0$ and, therefore, could not
explain why systems with ferromagnetic couplings like 
CeAl${}_3$\cite{andres} do not exhibit appreciable magnetic order.

The first and most popular attempt to implement in the Kondo
lattice this idea of explaining magnetic order as the result of
the competition between the RKKY interaction and the Kondo 
singlet formation was carried out by Doniach,\cite{doniach} who 
studied a one-dimensional analog of the KLM (the so-called `Kondo
necklace') in which the conduction electrons are represented
by pseudo spins. He took as variational antiferromagnetic
trial function, the direct product of local wave functions
consisting in a linear combination 
of the singlet and triplet states that
can be formed with the local spins and pseudo spins.
As a result, he found that magnetic order was destroyed for
$J>W$. Although, as he argued, this should only be understood as
a qualitative result, it is quite revealing since it
essentially says that, if the formation of Kondo singlets was
the mechanism responsible for the destruction of magnetic order,
this would take place when
$J$ is strong enough to overcome the exhaustion problem,
enabling the participation of the entire conduction band in the 
formation of Kondo singlets.

Another well-known approach to the problem of magnetic order in HF
systems is due to Coleman and Andrei\cite{coleman2},
who claimed that the Kondo scattering of low-energy conduction
electrons and local moments could stabilize the spin liquid against
an antiferromagnetic state which is supposed to have a rather
similar energy. This idea is still being investigated by some
authors.\cite{kikoin}

In contrast to these approaches, we have found in the last section
that when the RKKY interaction is explicitly extracted from the
Hamiltonian, the Kondo interaction is modified in such
a way that it no longer influences the ground state. Thus,
as we mentioned, the formation of a spin liquid has
to be attributed to the long-range character of the RKKY
interaction which produces strong frustration and enhances spin 
fluctuations in the system. Of course, this picture holds
only as long as the ground state $|\Phi\rangle$ remains
stable. In fact, we will later find that an electronic
instability appears for stronger couplings and, at that point,
the Kondo coupling between conduction electrons and local moments
will indeed influence the spin correlations.

It should be noted that, since $\tilde{H}_{\em RKKY}$ is invariant 
under the change of $J$ by $-J$, the discussion of magnetism is
(before the occurrence of the mentioned instability) the same for 
both ferro and antiferromagnetic couplings. Finally, we would
like to point out that other effects not contemplated in the
KLM like the existence of a small f-charge fluctuation could
also contribute to the stabilization of the spin liquid.

The approach presented in this paper, also leads to a physical
interpretation of the elementary excitations of the system
and their role in the various observable quantities which
is fundamentally different from the most popular interpretations.
We have seen in the previous section that the expression of
the KLH in terms of electron and hole
operators associated with the actual vacuum state
is inevitably accompanied by the appearance of
an explicit RKKY interaction which leads to the formation
of spin correlations. Thus, the physical picture of the system
at zero temperature motivated by our formalism is that of a
backgroung of spin correlations (a spin liquid) in which
two kind of modes can be excited: neutral spin modes
associated with deformations of the spin liquid and
charged modes corresponding to the excitation of electrons
and holes in the system.

The accurate determination of the state of the spin system
and its excitations would require finding the ground state
and spectrum of $\tilde{H}_{\em RKKY}$ which, as we mentioned,
is beyond the scope of this paper. Nevertheless, we can
still draw an important conlusion without having 
to solve this Hamiltonian: The thermal breakdown of the
spin correlations will produce an entropy increase
of the order of $N\ln 2$; since $\tilde{H}_{\em RKKY}$
is a very weak, second order interaction, the characteristic
temperature $T_{KL}$ associated with the spin correlations
will be very small and, therefore, this enormous entropy
increase will take place in a very small temperature interval,
leading to a huge specific heat of spin origin.

Thus, as we mentioned in the introduction, the formalism
developed in this paper supports the thesis of 
Kagan, Kikoin, and Prokof'ev,\cite{kagan} who argued that a strong
mass renormalization of the low-energy conduction
electrons cannot explain the high value and universal
character of the entropy measured at $T_{KL}$, concluding that
the heavy particles responsible for the low-temperature 
properties of the Kondo lattice should be {\em neutral fermions
of spin origin\/}, as opposed to the {\em charged\/} heavy fermions
of the slave-boson\cite{coleman,millis} and variational\cite{rice} 
approaches.

In the next two sections, we study the charged excitations
of the system.

\section{Charged excitations: Perturvative modes.}

Since we are interested in studying the zero-temperature charged
modes for a Kondo lattice in the {\em normal\/} state, we shall
assume that $|\Phi\rangle$ is a singlet spin liquid, namely,
\begin{equation}
{\bf S}|\Phi\rangle =0\, ,
\end{equation}
where ${\bf S}$ is the generator of spin rotations. The spin
rotational symmetry will be very useful to characterize the states
as eigenvectors of ${\bf S}^2$ and ${\bf S}^z$ with definite
eigenvalues. For a general state of the form $V|\Phi\rangle$,
where $V$ is a linear combination of products of 
$\tilde{e}^{\dag}_{{\bf p}\alpha}$, $\tilde{h}^{\dag}_{{\bf q}
\alpha}$, and $\tilde{s}_{l,i}$ operators, we can write
${\rm e}^{i{\bf \alpha}{\bf S}}V|\Phi\rangle =
{\rm e}^{i{\bf \alpha}{\bf S}}V
{\rm e}^{-i{\bf \alpha}{\bf S}}|\Phi\rangle$, 
which implies that the action of ${\bf S}$
on a state is given by 
\begin{equation}
{\bf S} V|\Phi\rangle =[{\bf S},V]|\Phi\rangle\, ,
\label{opers}
\end{equation}
We would like to emphasize that, since the transformation 
(\ref{ttrans}) preserves the spin symmetry,
the commutation rules of ${\bf S}$ with the initial and
transformed operators are exactly the same.
In particular, Eq.\ (\ref{opers}) implies that the states 
$\tilde{e}^{\dag}_{{\bf p}\alpha}|\Phi\rangle$ and
$\tilde{s}_{l,{\bf k}}|\Phi\rangle$ form, respectively, an
$s=1/2$ charged doublet and an $s=1$ triplet of spin fluctuations.
The fact that $S^z \tilde{s}_{l,{\bf k}}|\Phi\rangle=l
\tilde{s}_{l,{\bf k}}|\Phi\rangle$ justifies the notation
introduced in Eq.\ (\ref{smalls}) for the $s$ operators.

Aside from the contributions of spin correlations, 
the dispersion relation corresponding to the states
$\tilde{e}^{\dag}_{{\bf p}\alpha}|\Phi\rangle$ and
$\tilde{h}^{\dag}_{{\bf q}\alpha}|\Phi\rangle$ is given
by Eq.\ (\ref{disprel0}) and contains a gap of magnitude 
$2\tilde{E}(0)=\tilde{\eta}+3J^2/16$. 
This statement, however, does not have any physical significance 
because these states are not the charged excitations of the system.

For small $J$, the actual charged modes are expected to be
slight modifications of the states 
$\tilde{e}^{\dag}_{{\bf p}\alpha}|\Phi\rangle$ and
$\tilde{h}^{\dag}_{{\bf q}\alpha}|\Phi\rangle$ and, consequently,
we will refer to them as {\em perturbative modes\/}. However,
as we argued in the introduction in relation to the Nozi\`eres
exhaustion problem,
we should also consider the possibility that the Kondo interaction
would produce a strong correlation between low-energy electrons
and spin waves, leading to the formation of collective states.
These {\em strongly correlated modes\/} will be investigated in 
the next section.

When we determined ${\rm e}^{\tilde{T}}$ in Section II, we argued that
the transformation ${\rm e}^{\tilde{T}_2}$ [Eq.\ (\ref{2t})] did not 
change $|\Phi\rangle$ and, therefore, could be set equal to
the identity. Nevertheless, since ${\rm e}^{\tilde{T}_2}$ {\em does\/}
change the basic operators $\tilde{c}_{{\bf k}\alpha}$ and 
$\tilde{s}_{l,{\bf k}}$, we can use the degrees of freedom
associated to this transformation to get a new set of canonical
operators $\hat{c}_{{\bf k}\alpha}$ and $\hat{s}_{l,{\bf k}}$ 
which still satisfy $\hat{e}_{{\bf p}\alpha}|\Phi\rangle=
\hat{h}_{{\bf q}\alpha}|\Phi\rangle=0$ and which are optimal in
the sense that $\hat{e}^{\dag}_{{\bf p}\alpha}|\Phi\rangle$ and
$\hat{h}^{\dag}_{{\bf q}\alpha}|\Phi\rangle$ are the best possible
approximations to the actual charged modes.

In the next section, we will calculate this second transformation
($\hat{T}$), and will find that $\hat{c}_{{\bf k}\alpha}$, 
$\hat{s}_{l,{\bf k}}$
are slight deformations of the tilded operators.
Since $\hat{c}_{{\bf k}\alpha}$, $\hat{s}_{l,{\bf k}}$ 
are optimal operators, they should be the basis to construct the 
actual charged modes.  In fact, they will be used in the calculation
of the strongly correlated modes. To construct the perturbative modes,
however, we can also use the tilded operators because the
composition of two small deformations is still a small deformation.
In fact, the direct calculation with the tilded 
operators is more reliable because we avoid the inevitable 
approximations associated with carrying out a second transformation.

Since no symmetry is broken and no bound states are created at this
point, there is no physical reason for the appearance of a gap
in the charged spectrum. In fact, we will now calculate the
dispersion relation for the perturbative modes to leading order
in $J$ neglecting the effect of the spin correlations,
showing how the gap is indeed closed in this order.
The perturbative modes must have $s=1/2$. Due to the spin rotational
symmetry, the two states in a spin doublet will have the same
energy and, therefore, we can select, for instance, the state
with $s^z=1/2$ to carry out our calculations. Moreover, since
the situation considered is symmetric with respect to electrons
and holes, we only need to calculate the spectrum corresponding,
for example, to the electrons. It can be seen, using Eq.\ 
(\ref{opers}), that a small deformation of the state
$\tilde{e}^{\dag}_{{\bf p}\uparrow}|\Phi\rangle$ having negative
charge, $s^z=s=1/2$, and wave vector ${\bf p}$, has the structure
\begin{equation}
|PE_{{\bf p}\uparrow}\rangle=\alpha({\bf p})\left[
\tilde{e}^{\dag}_{{\bf p}\uparrow}
-\frac{J}{4N^{1/2}}\beta_{\bf p}({\bf p'})
(\tilde{e}^{\dag}_{{\bf p'}\uparrow}\tilde{s}_{0,{\bf p}-
{\bf p'}}+\sqrt{2}\tilde{e}^{\dag}_{{\bf p'}\downarrow}
\tilde{s}_{1,{\bf p}-{\bf p'}})\right]|\Phi\rangle\, ,
\label{anspe}
\end{equation}
where $PE$ stands for `perturbative electron',
$\beta_{\bf p}({\bf p'})$ is an arbitrary variational function, and
$\alpha({\bf p})$ is a normalization factor given by
\begin{equation}
\alpha({\bf p})=\left[1+\frac{3J^2}{16\Gamma}\int d{\bf p'}
\beta^{\star}_{\bf p}({\bf p'})\beta_{\bf p}({\bf p'})\right]^{-1/2}.
\label{adepe}
\end{equation}

The function $\beta_{\bf p}({\bf p'})$ should be selected to turn
the state (\ref{anspe}) into an energy eigenstate. This condition
can be implemented by demanding that $\beta_{\bf p}({\bf p'})$ be
an extreme of the functional
$E([\beta],{\bf p})\equiv \langle PE_{{\bf p}\uparrow}|H|
PE_{{\bf p}\uparrow}\rangle$.

Obviously, to calculate this quantity, we will use the expression
of $H$ in terms of the tilded operators [Eq.\ (\ref{1stth})].
The scalar product in this equation can be developed with Eqs.\
($\tilde{\ref{anticomrel}}$), ($\tilde{\ref{mtable}}$), and
($\tilde{\ref{adjoints}}$) into an expression in which only
scalar products of the form $\langle\Phi|\tilde{s}_{l_1,i_1}\ldots
\tilde{s}_{l_m,i_m}|\Phi\rangle$ with $i_1,\ldots,i_m$ corresponding
to different lattice sites, leave to be determined. These are
precisely the spin correlation functions, and depend on the
magnetic structure of $|\Phi\rangle$ arising from the RKKY
interaction. As we mentioned, the calculation of these correlations
is beyond the scope of this paper and, therefore, its effect on
the charged modes of the system will be neglected. Thus, we will
calculate these modes assuming a disordered 
spin lattice, which means that all the spin 
correlation functions vanish, enabling a simple computation
of the scalar products. Since the constant $\tilde{C}$ in Eq.\ 
(\ref{1stth}) coincides, in this case, with the ground state
energy, we will remove it from $H$ assigning, in this way,
zero energy to $|\Phi\rangle$. This means that the energy of
every state is measured with respect to the vacuum state, and represents,
therefore, its {\em excitation\/} energy. Obviously, this energy has to be
positive; in fact, the existence of a state with negative energy, would
imply an instabily of $|\Phi\rangle$ and a transition to a 
new phase.

With all these conditions taken into account, the calculation of 
$E([\beta],{\bf p})$ yields
\begin{eqnarray}
E(&&[\beta],{\bf p})=\tilde{E}(\varepsilon_{\bf p})-\frac{3J^2}{16}
\alpha^2({\bf p})\left\{ \frac{\tilde{\alpha}(\varepsilon_{\bf p})}
{\Gamma}\int\! d{\bf p'}\tilde{\alpha}(\varepsilon_{\bf p'})
\left[\beta^{\star}_{\bf p}({\bf p'})+\beta_{\bf p}({\bf p'})\right]\right.
\label{eofbp} \\
&&\left. -\frac{1}{\Gamma}\int\! d{\bf p'}
\left[\tilde{E}(\varepsilon_{\bf p'})-\tilde{E}
(\varepsilon_{\bf p})\right]\beta^{\star}_{\bf p}({\bf p'})
\beta_{\bf p}({\bf p'})+\frac{J}{2\Gamma^2}\int\! d{\bf p_1}d{\bf p_2}
\tilde{\alpha}(\varepsilon_{\bf p_1})\tilde{\alpha}
(\varepsilon_{\bf p_2})\beta^{\star}_{\bf p}({\bf p_1})\beta_{\bf p}
({\bf p_2})\right\} ,
\nonumber
\end{eqnarray}
where $\tilde{\alpha}(\varepsilon)$ and $\tilde{E}(\varepsilon)$
are given by Eqs.\ (\ref{alpha}) and (\ref{disprel0}) respectively.
The last term of this equation is of higher order in $J$ than
the others, and can be neglected.

If the gap is indeed closed, $E(0)\equiv E([\beta],{\bf p}_F)$ 
(where ${\bf p}_F$ is a wave vector of the Fermi surface) must
vanish. This quantity can be written as $E(0)=\tilde{E}(0)-\Lambda$, 
where $\Lambda$ represents the energy decrease due to the Kondo 
interaction, and is given by
\begin{equation}
\Lambda=\frac{3J^2\alpha^2({\bf p}_F)}{16\Gamma}\int\! d{\bf p}
\left\{\tilde{\alpha}(0)\tilde{\alpha}(\varepsilon_{\bf p})
\left[\beta^{\star}({\bf p})+\beta({\bf p})\right]-\left[
\tilde{E}(\varepsilon_{\bf p})-\tilde{E}(0)\right]
\beta^{\star}({\bf p})\beta({\bf p})\right\},
\label{lamb}
\end{equation}
where $\beta({\bf p})\equiv \beta_{{\bf p}_F}({\bf p})$.
Functionally differentiating 
with respect to $\beta({\bf p})$ in this equation, we find that the 
condition of extreme of $E(0)$ is given by
\begin{equation}
\beta({\bf p})=\frac{\tilde{\alpha}(0)\tilde{\alpha}(\varepsilon_{\bf 
p})}{\tilde{E}(\varepsilon_{\bf p})-\tilde{E}(0)+\Lambda}\, ,
\label{bpsol}
\end{equation}
and substituting this expression in Eq.\ (\ref{lamb}), we obtain the
following equation for $\Lambda$:
\begin{equation}
\Lambda=\frac{3J^2}{16}\tilde{\alpha}^2(0)\int_0^{1/2}\!
d\varepsilon\frac{\tilde{\alpha}^2(\varepsilon)}
{\tilde{E}(\varepsilon)-\tilde{E}(0)+\Lambda}\, .
\label{lambeq}
\end{equation}

As we already mentioned, $\tilde{\alpha}(\varepsilon)$ can
be substituted, to leading order in J, by 1. Also,
since $\tilde{E}(\varepsilon)-\tilde{E}(0)$ vanishes at 
$\varepsilon=0$, we can approximate it by $\varepsilon$.
Thus, with the approximations
\begin{equation}
\tilde{\alpha}(\varepsilon)\sim 1\;\; , \;\;
\tilde{E}(\varepsilon)\sim\tilde{E}(0)+\varepsilon\, ,
\label{tildeapp}
\end{equation}
Eq.\ (\ref{lambeq}) is turned into 
\begin{equation}
\Lambda=\frac{3J^2}{16}
\ln\frac{1}{2\Lambda}\, ,
\label{applambda}
\end{equation}
which implies that
$\Lambda=\tilde{\eta}/2+O(J^2)$. Since $\tilde{E}(0)=\tilde{\eta}/2 
+3J^2/32$, this proves the cancellation of the leading contributions 
to $E(0)$. We will not verify the cancellation of the $J^2$ terms
because, since these are lower order contributions, to
accurately calculate them, we should have to control all the 
approximations that have enabled us to carry out our computations 
analytically. Since, as we already mentioned, there 
is no physical reason for the formation of a gap, we will assume
that the $J^2$ contributions cancel as well.

Finally, we would like to point out that if the spin correlations 
were considered, we should include in $H$ the terms
$\tilde{e}^{\dag}\tilde{e}\tilde{s}_i\tilde{s}_j$ 
and $\tilde{h}^{\dag}\tilde{h}\tilde{s}_i\tilde{s}_j$ ($i\neq j$)
that we mentioned at the end of Section II because, in this
case, they give leading contributions to $E(0)$ which compete with
those steming from the kinetic and Kondo terms.

\section {Second transformation, strongly correlated charged modes,
and vacuum instability}

In this section, we shall investigate the possible formation of
strongly correlated charged modes. As we already argued in
Section IV (and will become even more clear at the end of this
section), these modes should be written in terms of optimal
fermionic operators representing the best possible approximations
to the actual (perturbative) charged excitations. Thus, we will
start by carrying out a second unitary transformation to determine
these operators, and will find the expression of the Hamiltonian
in terms of them. Then, with these building blocks, we will
construct the strongly correlated charged modes and will discuss
their nature in the case of ferro and antiferromagnetic couplings.
Finally, we will show that, for sufficiently strong couplings, these 
modes condensate in the vacuum state leading to an instability.

In analogy with the notation ($\tilde{\rm x}$), ($\hat{\rm x}$) will
denote the equation equivalent to (x) for the optimal operators
$\hat{c}_{{\bf k}\alpha}$, $\hat{s}_{l,{\bf k}}$. Since we want to 
continue preserving all the symmetries of $H$, the second unitary 
transformation $\hat{T}$ that we will now determine must have
the structure ($\hat{\ref{ttrans}}$), and the relations
between the tilded and hatted operators will be given by Eqs.\
($\hat{\ref{cpno}}$)--($\hat{\ref{s1no}}$) with 
$\hat{\alpha}({\bf k})$ equal, for the moment, to one. To ensure that
$\hat{e}_{{\bf p}\alpha}|\Phi\rangle=\hat{h}_{{\bf q}\alpha}|\Phi
\rangle=0$, ${\rm e}^{\hat{T}}$ must leave $|\Phi\rangle$ invariant,
which requires that $\hat{T}({\bf p},{\bf q})=\hat{T}({\bf q},{\bf p})
=0$.

We shall take as starting point in the following discussion,
the expression (\ref{1stth}) of $H$ in terms of the tilded
operators with the approximations (\ref{tildeapp}). The
use of these approximations in the calculation of quantities
in which the dominant contributions cancel will be justified.
The main effect of the Kondo interaction is, obviously, to
renormalize the charged excitations. Since $\hat{e}^{\dag}_{{\bf p}
\alpha}|\Phi\rangle$ and $\hat{h}^{\dag}_{{\bf q}\alpha}|\Phi\rangle$
are optimal approximations to the perturbative charged modes, they
will include these renormalization effects and, therefore, they will
be subject only to a small residual Kondo interaction. Thus, when we
write $H$ in terms of the hatted operators, the bulk of the Kondo
interaction should disappear. In analogy with equation (\ref{fot}),
the condition for the cancellation of the terms $\hat{e}^{\dag}
\hat{e}\hat{s}$ and $\hat{h}^{\dag}\hat{h}\hat{s}$ to leading order
in $J$ is given by
\begin{equation}
\hat{T}({\bf p},{\bf p'})=\frac{1}{\varepsilon_{\bf p}-
\varepsilon_{\bf p'}}\;\;,\;\;
\hat{T}({\bf q},{\bf q'})=\frac{1}{\varepsilon_{\bf q}-
\varepsilon_{\bf q'}}\, .
\label{tpptqq}
\end{equation}

Again, it can be shown that the singularities of these expressions
at the points where $\varepsilon_{\bf k}=\varepsilon_{\bf k'}$ lead to a
loss of unitarity. Thus, we need, once more, to introduce a regulator
to tame the singularities. In contast to $\varepsilon_{\bf p}-
\varepsilon_{\bf q}$ [Eq.\ (\ref{tpq0})] which is always positive,
$\varepsilon_{\bf p}-\varepsilon_{\bf p'}$ and $\varepsilon_{\bf q}
-\varepsilon_{\bf q'}$ can have positive and negative values and,
therefore, to avoid the vanishing of the denominators of
Eq.\ (\ref{tpptqq}), we should add to them a small {\em imaginary\/} 
quantity. Thus, the regulated forms of $\hat{T}({\bf p},{\bf p'})$
and $\hat{T}({\bf q},{\bf q'})$ can be written as
\begin{equation}
\hat{T}({\bf p},{\bf p'})=\frac{1}{2}\left[ \frac{1}
{\varepsilon_{\bf p}-\varepsilon_{\bf p'}+i\hat{\eta}}
+\frac{1}{\varepsilon_{\bf p}-\varepsilon_{\bf p'}-i\hat{\eta}}
\right]=\frac{\varepsilon_{\bf p}-\varepsilon_{\bf p'}}
{(\varepsilon_{\bf p}-\varepsilon_{\bf p'})^2+\hat{\eta}^2}\,
\end{equation}
\begin{equation}
\hat{T}({\bf q},{\bf q'})=\frac{\varepsilon_{\bf q}-
\varepsilon_{\bf q'}}{(\varepsilon_{\bf q}-\varepsilon_{\bf q'})^2
+\hat{\eta}^2}\, ,
\end{equation}
where, to re-establish unitarity for small couplings, the real constant 
$\hat{\eta}$ should vanish more slowly than $J^2$ as $J\to 0$.
Again, as we did with the first transformation in Section II, to
attain unitarity to the next-to-leading order in $J$, we should
introduce in Eqs.\ ($\hat{\ref{cpno}}$) and ($\hat{\ref{cno}}$)
the factor $\hat{\alpha}({\bf k})$ given by
\begin{equation}
\hat{\alpha}({\bf k})=\left[ 1+\frac{3J^2}{16N}\sum_{\bf k'}
\hat{T}^2({\bf k'},{\bf k})\right]^{-1/2}=
\left[1+\frac{3J^2}{16}\int_0^{1/2}\!dx\frac{(|\varepsilon_{\bf k}|
-x)^2}{\left[(|\varepsilon_{\bf k}|-x)^2+\hat{\eta}^2\right]^2}
\right]^{-1/2}.
\label{alphahat}
\end{equation}

To completely specify this second transformation, we just need
to determine $\hat{\eta}$. We saw in the last section that the
Kondo interaction lowered the energy of the
charged states around the Fermi level; in fact, the equations
(\ref{bpsol}) and (\ref{lambeq}) which determined the state 
$|PE_{{\bf p}_F\uparrow}\rangle$, are nothing but the conditions
for minimizing the energy of this state. Since 
$\hat{e}^{\dag}_{{\bf p}_F\uparrow}|\Phi\rangle$ must be the
best possible approximation to $|PE_{{\bf p}_F\uparrow}\rangle$,
we will determine $\hat{\eta}$ by demanding that its energy 
be minimized. 

If the spin correlations are disregarded,
the energy of the states $\hat{e}^{\dag}_{{\bf p}\alpha}|\Phi\rangle$
is given by the coefficient $\hat{E}({\bf p})$ of the kinetic term
($\sum_{{\bf p}\alpha}\hat{E}({\bf p})\hat{e}^{\dag}_{{\bf p}\alpha}
\hat{e}_{{\bf p}\alpha}$) of $H$ expressed in terms of the hatted
operators. In particular, $\hat{e}^{\dag}_{{\bf p}_F\uparrow}|
\Phi\rangle$ has the energy $\hat{E}(0)=\tilde{E}(0)-\hat{\Lambda}$
where $\hat{\Lambda}$ is given, with the approximations 
(\ref{tildeapp}) for $\tilde{\alpha}(\varepsilon)$ and
$\tilde{E}(\varepsilon)$, by
\begin{equation}
\hat{\Lambda}=\frac{3J^2}{16}\int_0^{1/2}\!d\varepsilon\left(
2\hat{\alpha}(0)\hat{\alpha}(\varepsilon)-\hat{\alpha}^2
(\varepsilon)\frac{\varepsilon^2}{\varepsilon^2+\hat{\eta}^2}\right)
\frac{\varepsilon}{\varepsilon^2+\hat{\eta}^2}\, ,
\end{equation}
where the function $\hat{\alpha}(\varepsilon)$ is given by
Eq.\ (\ref{alphahat}). After some calculations, it can be shown
that the condition of extreme of $\hat{\Lambda}$ ($d\hat{\Lambda}/
d\hat{\eta}=0$) leads to the following equation for $\hat{\eta}$:
\begin{equation}
\hat{\eta}=\frac{3\pi J^2}{64}\ln\frac{1}{2\hat{\eta}}\, ,
\label{etahat}
\end{equation}
and that the corresponding (maximum) value of $\hat{\Lambda}$ is given,
to order $J^2$, by 
\begin{equation}
\hat{\Lambda}=\frac{3J^2}{16}\left(\ln\frac{1}
{2\hat{\eta}}-\frac{1}{2}\right) .
\label{lamhatapp}
\end{equation}

$\hat{E}(0)$ can also be written as $\hat{E}(0)=E(0)+\Lambda-
\hat{\Lambda}=\Lambda-\hat{\Lambda}$. Since the dominant parts of
$\Lambda$ and $\hat{\Lambda}$ cancel in this equation, $\hat{E}(0)$
will be proportional to $J^2$ and it seems that, to calculate the 
coefficient of proportionality, an accurate computation controlling all 
the approximations [like (\ref{tildeapp})] would be needed.
However, although these approximations modify
the order $J^2$ of the energy decreases $\Lambda$ and $\hat{\Lambda}$ 
corresponding to the states $|PE_{{\bf p}_F\uparrow}\rangle$ and 
$\hat{e}^{\dag}_{{\bf p}_F\uparrow}|\Phi\rangle$, their effect on these 
two states must be very similar and, therefore, they should
not essentially change $\hat{E}(0)=\Lambda-\hat{\Lambda}$. Thus, 
by substituting in this equation the values of $\Lambda$ and
$\hat{\Lambda}$ calculated with the same approximations in
Eqs.\ (\ref{applambda}) and (\ref{lamhatapp}), 
we should get an estimate of $\hat{E}(0)$. 
The result obtained in this way is 
\begin{equation}
\hat{E}(0)=\Lambda-\hat{\Lambda}=\frac{3J^2}{16}\left(\ln\frac{\pi}{4}
+\frac{1}{2}\right)=0.048J^2\, .
\end{equation}

Having determined $\hat{\eta}$ [Eq.\ (\ref{etahat})], the relations
($\hat{\ref{cpno}}$)--($\hat{\ref{s1no}}$) between the tilded and
hatted operators are completely specified. In analogy
with what we did in Section II with the first transformation,
by substituting these relations in Eq.\ (\ref{1stth}), normal-ordering 
the hatted operators corresponding to electrons and holes, and
contracting the operators $\hat{s}_{l,i}$ at the same lattice
sites, we obtain the expression of $H$ in terms of
the optimal coordinates $\hat{e}_{{\bf p}\alpha}$,
$\hat{h}_{{\bf q}\alpha}$, and $\hat{s}_{l,{\bf k}}$.
By carefully examining all the terms generated in this process,
we can discriminate the ones which are dominant for small $J$.
After this analysis, we find that the new expression of $H$ is
essentially given by
\begin{equation}
H=\tilde{C}+\hat{H}_{\em kinetic}+\hat{H}_{\em Kondo}+
\hat{H}_{\em RKKY}\, .
\label{2ndth}
\end{equation}
where
\begin{eqnarray}
\hat{H}_{\em kinetic}&=&\sum_{{\bf p},\alpha}
\hat{E}(\varepsilon_{\bf p})\hat{e}^{\dag}_{{\bf p}\alpha}
\hat{e}_{{\bf p}\alpha}+\sum_{{\bf q},\alpha}
\hat{E}(-\varepsilon_{\bf q})\hat{h}^{\dag}_{{\bf q}\alpha}
\hat{h}_{{\bf q}\alpha}\;\;{\rm with}\;\; \hat{E}(\varepsilon)\sim
\hat{E}(0)+\varepsilon\, , \\
\hat{H}_{\em Kondo}&=&\frac{\hat{\eta}^2}{(\varepsilon_{\bf p}-
\varepsilon_{\bf p'})^2+\hat{\eta}^2}\times(\hat{\ref{epest}})
+\frac{\hat{\eta}^2}{(\varepsilon_{\bf q}-\varepsilon_{\bf q'})^2
+\hat{\eta}^2}\times(\hat{\ref{hphst}})\, , \label{kondohat} \\
\hat{H}_{\em RKKY}&=&\frac{1}{2}\sum_{i\neq j}J_{\em RKKY}
({\bf R}_i -{\bf R}_j)\hat{\bf S}_i\hat{\bf S}_j\, .
\end{eqnarray}

Thus, as expected, the main effects of the second transformation
are to strongly reduce the gap in the kinetic term and to substitute
the Kondo coupling by a residual Kondo interaction which is effective
only between electrons or holes having very similar energies. In contrast
to what happened with the first transformation, the technical reason
for the appearance of this residual Kondo interaction is that, in this
case, the contributions of higher order terms are not important and
cannot cancel the Kondo coupling in the regions where 
$\varepsilon_{\bf k}\sim\varepsilon_{\bf k'}$.

Armed with this (optimal) expression of $H$, we will now look for
strongly correlated charged states, namely, we will investigate
the possibility that an electron or a hole combines with a wave of spin
fluctuations forming a collective state. Again, since the situation is 
the same for positively and negatively charged modes, we will only need 
to investigate, for example, the last ones. The structure of such a 
correlated state having negative charge, wave vector ${\bf k}$ and 
$s=s^z=1/2$ must be
\begin{equation}
|SCE_{{\bf k},1/2}\rangle=N^{-1/2}A B({\bf p})\left( 
\hat{e}^{\dag}_{{\bf p}\uparrow}\hat{s}_{0,{\bf k}-{\bf p}}+
\sqrt{2}\hat{e}^{\dag}_{{\bf p}\downarrow}
\hat{s}_{1,{\bf k}-{\bf p}}\right)|\Phi\rangle\, ,
\label{sce}
\end{equation}
where $SCE$ stands for `strongly correlated electron', $B({\bf p})$
is a variational function, and $A$ is a normalization
factor given by
\begin{equation}
A=\left[ \frac{3}{\Gamma}\int\! d{\bf p} B^{\star}({\bf p})
B({\bf p})\right]^{-1/2} .
\end{equation}

Neglecting the spin correlations, 
the energy of this state, measured with respect to the ground
state energy, is given by
\begin{equation}
E_{SC,1/2}[B]=A^2\left[\frac{3}{\Gamma}\int\! d{\bf p}\hat{E}
(\varepsilon_{\bf p}) B^{\star}({\bf p})B({\bf p})-\frac{3}
{2{\Gamma}^2}\int\! d{\bf p}d{\bf p'}B^{\star}({\bf p'})
\hat{J}(\varepsilon_{\bf p'},\varepsilon_{\bf p})
B({\bf p})\right],
\label{esc}
\end{equation}
where $\hat{J}(\varepsilon_1,\varepsilon_2)=J\hat{\eta}^2/
[(\varepsilon_1-\varepsilon_2)^2+\hat{\eta}^2]$ is the residual
Kondo coupling of Eq.\ (\ref{kondohat}).

Obviously, the variational function $B({\bf p})$ will be determined
by demanding that $E_{SC,1/2}$ is minimized. This condition is given by
\begin{equation}
(\hat{E}(\varepsilon_{\bf p})-E_{SC,1/2})B({\bf p})-\frac{1}{2\Gamma}
\int\!d{\bf p'}\hat{J}(\varepsilon_{\bf p},\varepsilon_{\bf p'})
B({\bf p'})=0\, .
\label{bofpeq}	
\end{equation}

From Eqs.\ (\ref{esc}) and (\ref{bofpeq}), we should be able to
calculate $E_{SC,1/2}$ and $B({\bf p})$. It should be noted, however,
that, to minimize $E_{SC,1/2}$, the kinetic part in Eq.\ (\ref{esc})
must be kept small, which requires that $B({\bf p})$ have all its 
weight concentrated in a very narrow layer around the Fermi level.
Since the residual Kondo coupling is essentially constant and equal
to $J$ in this region, we will make the approximation of substituting
in Eq.\ (\ref{bofpeq}) the function $\hat{J}(\varepsilon_{\bf p},
\varepsilon_{\bf p'})$ by $J$, and will prove later on that this is
indeed a valid approximation. Thus, from Eq.\ (\ref{bofpeq}) we have
\begin{equation}
B({\bf p})=\frac{Z}{\hat{E}(\varepsilon_{\bf p})-E_{SC,1/2}}\;\; , \;\;
Z=\frac{J}{2\Gamma}\int\! d{\bf p}B({\bf p})\, .
\end{equation}
Substituting the first of these equations into the second one, we get
the following condition for $E_{SC,1/2}$:
\begin{equation}
1=\frac{J}{2}\int_0^{1/2}\frac{d\varepsilon}{\varepsilon +
\hat{E}(0)-E_{SC,1/2}}\, .
\end{equation}
This equation has a solution only for antiferromagnetic couplings
($J>0$), and it is given by
\begin{equation}
E_{SC,1/2}=\hat{E}(0)-\Omega_{1/2}\;\; , \;\; \Omega_{1/2}=
\frac{{\rm e}^{-2/J}}{2}\, .
\label{omega12}
\end{equation}

For ferromagnetic couplings, the local moments and the spin of the
conduction electrons will tend to align parallel to each other and,
if a collective state is formed, it should have the structure
\begin{equation}
|SCE_{{\bf k},3/2}\rangle=N^{-1/2}CD({\bf p})\hat{e}^{\dag}_{{\bf p}
\uparrow}\hat{s}_{1,{\bf k}-{\bf p}}|\Phi\rangle\, ,
\end{equation}
which is the general form of a state composed of a charge carrier
and a spin wave having wave vector ${\bf k}$, negative charge, and
$s=s^z=3/2$. Again, $D({\bf p})$ is a variational function and $C$
a normalization factor. The energy of this state in the absence
of spin correlations is given by
\begin{equation}
E_{SC,3/2}[D]=C^2\left[\frac{1}{\Gamma}\int\! d{\bf p}\hat{E}
(\varepsilon_{\bf p}) D^{\star}({\bf p})D({\bf p})+\frac{1}
{4{\Gamma}^2}\int\! d{\bf p}d{\bf p'}D^{\star}({\bf p'})
\hat{J}(\varepsilon_{\bf p'},\varepsilon_{\bf p})
D({\bf p})\right].
\end{equation}
Following the same steps used to estimate $E_{SC,1/2}$, we arrive
to an equation for $E_{SC,3/2}$,
\begin{equation}
1=\frac{-J}{4}\int_0^{1/2}\frac{d\varepsilon}{\varepsilon +
\hat{E}(0)-E_{SC,3/2}}\, ,
\end{equation}
which has solution only for ferromagnetic couplings and it is given
by
\begin{equation}
E_{SC,3/2}=\hat{E}(0)-\Omega_{3/2}\;\; , \;\; \Omega_{3/2}=
\frac{{\rm e}^{4/J}}{2}\, .
\label{omega32}
\end{equation}

Thus, summarizing, we have found that in ferro and antiferromagnetic
Kondo lattices there is a tendency to form collective states of spin
$3/2$ and $1/2$ respectively, and that, for couplings with the same
strength, the formation of these states is much more favored in the
antiferromagnetic case ($\Omega_{1/2}\gg \Omega_{3/2}$).

If $\Omega$ stands for either $\Omega_{1/2}$ or $\Omega_{3/2}$, the
probability of finding in a SC mode an electron with wave vector
${\bf p}$ in the layer $\varepsilon\leq\varepsilon_{\bf p}\leq
\varepsilon+d\varepsilon$ is $P(\varepsilon)d\varepsilon$, where
$P(\varepsilon)\simeq\Omega/(\varepsilon+\Omega)^2$. This means,
for instance, that 50\% of the electrons in these modes have their
wave vectors in the layer $0\leq\varepsilon_{\bf p}\leq\Omega$.
Thus, if $\Omega\ll\hat{E}(0)$, the SC modes cannot form because
it is energetically more favorable for the electrons in this layer
to remain in the PE modes. However, if $\Omega$ is smaller but
of the order of $\hat{E}(0)$, the low-energy electrons are
energetically allowed to participate in the formation of both
PE and SCE and this will result in a very strong renormalization
of the electron (and hole) masses around the Fermi level. Finally,
if $\Omega>\hat{E}(0)$, the energy of the SCE will become negative,
which means that these states condensate in the ground state
producing an instability and the transition to a new phase.

It should be remembered that, to simplify the calculations,
we took a constant density of states $D=1$. However, in a real band
$D(\varepsilon)$ vanishes at the edges and it is enhanced 
in the middle of the band (Fermi level). 
Since, as we have already mentioned, the SCE are composed of 
very soft electrons, only the density of states at the Fermi level
($D_F$) plays a role in their formation. Thus, more realistic
expressions of $\Omega$ are
\begin{equation}
\Omega_{1/2}=\frac{{\rm e}^{-2/JD_F}}{2}\;\; , \;\;
\Omega_{3/2}=\frac{{\rm e}^{4/JD_F}}{2}\, .
\end{equation}
Since the value of $\Omega$ depends very strongly on the exponents 
of these expressions, it is very important to use the 
correct value of $D_F$ when we try to estimate 
$J_c$, namely, the value of $J$ at the phase transition point
($\Omega=\hat{E}(0)$). For example, in the antiferromagnetic case we 
have (in units of $W$) $J_c(D_F=1.3)=0.34$, $J_c(D_F=1.5)=0.27$, 
and $J_c(D_F=2)=0.17$.

At this point, we would like to show, as we promised, that the 
approximation $\hat{J}(\varepsilon_{\bf p},\varepsilon_{\bf p'})
\sim J$ used to determine $E_{SC}$ is indeed justified: The average
value of the residual Kondo coupling $\hat{J}(\varepsilon_1,
\varepsilon_2)$ in a SCE state is given by
\begin{equation}
\langle\hat{J}\rangle =\int_0^{1/2}\int_0^{1/2}\! d\varepsilon_1
d\varepsilon_2 P(\varepsilon_2)\hat{J}(\varepsilon_1,\varepsilon_2)
P(\varepsilon_1)\, .
\end{equation}
Substituting the expressions for $P(\varepsilon)$ and $\hat{J}
(\varepsilon_1,\varepsilon_2)$, this equation can be developed into
\begin{equation}
\frac{\langle\hat{J}\rangle}{J}=1-2\int_0^{\frac{1}{2\Omega}}\!
\frac{dx}{x^2+\hat{\eta}^2/{\Omega}^2}\left(\frac{x+2}{x+1}-
\frac{2}{x}\ln(x+1)\right).
\end{equation}
Since the second term of this equation is small when 
$\hat{\eta}/\Omega\gg 1$ (the actual physical case), 
$\langle\hat{J}\rangle\sim J$, and our approximation is justified. 
For example, in the
antiferromagnetic case with $D_F=1.5$ we have, at $J=J_c$,
$\hat{\eta}/\Omega=8.6$ and $\langle\hat{J}\rangle=0.81 J$. 

It is interesting to note that in our calculation of the SCE
starting from the Hamiltonian (\ref{1stth}), we have reduced the
minimum energy $\tilde{E}(0)$ of the states 
$\tilde{e}^{\dag}_{{\bf p}\alpha}|\Phi\rangle$ in two steps: First
finding optimal operators $\hat{e}^{\dag}_{{\bf p}\alpha}$ which
strongly reduced the minimum energy from $\tilde{E}(0)$ to
$\hat{E}(0)$ using the high-energy tail of the Kondo coupling
and, second, constructing with these operators SC modes which
further reduced the energy using the low-energy part of the Kondo
coupling. This makes it very clear why we needed to find the
optimal operators to construct the SCE: If we would have built
these modes with the operators $\tilde{e}^{\dag}_{{\bf p}\alpha}$
and $\tilde{s}_{l,{\bf k}}$, we would not have cacelled the
high-energy tail of the Kondo coupling, and the resulting states
could not be considered as proper states because they would continue
interacting strongly with the lattice of local moments.

As we mentioned, we have always neglected the influence of the spin
correlations on the charged modes. This influence is actually
expected to be important in the quantitative description of the
SC modes. At first sight, it seems that these correlations would
tend to inhibit the formations of SCE because an additional energy 
(proportional to $J^2$) would be needed to excite the spin waves
present in these modes; however, the changes in the scalar
product introduced by the spin correlations would also 
modify the contribution
of the Kondo interaction to the energy and the overall effect is
hard to predict. 

The most important challenge posed by this article is the description
and characterization of the new phase. Obviously, since we know
experimentally that the instabilities in HF systems are associated
with antiferromagnetim and superconductivity, it is important to
elucidate whether or not the instability that we have described
is associated with the onset of superconductivity. In any case,
it is clear that the condensation in the ground state of collective
charged states will dramatically change
the diamagnetic and transport properties
of the system. It should also be noted that, since in these condensated
states, spin waves strongly couple to conduction electrons, the transition
to the new phase will be accompanied by changes in the spin correlations.
Thus, since we have attributed the high values of the low-temperature
specific heat to the thermal breakdown of soft spin correlations, an
important change in the specific heat is expected to occur at
the transition point. In fact, since the state of the spin system affects
the SC modes and the condensation of these modes alters the spin system,
an accurate calculation of the transition point would require a 
self-consistent calculation involving the two systems.

\section{Summary and conclusions}

In this article, we have assumed that the KLM incorporates the
essential physics of HF systems and, investigating this model
at zero temperature and half filling with a `variational perturbation 
theory', we have obtained the following physical picture:

\begin{itemize}
\item The non-interacting system consists in an electronic Fermi 
sea with electrons and holes as elementary excitations, plus a 
completely degenerate (random) spin lattice.
\item For small couplings, the charged spectrum remains basically
unchanged but an induced RKKY interaction  breaks the degeneracy of the
spin lattice. 
\item At this stage, the RKKY interaction is the only one responsible
for the spin dynamics. This very long-range, oscillating interaction
implies strong frustration and the enhancement of spin fluctuations,
which should explain the formation of a spin liquid or the appearance of a
weak magnetic order depending on the crystal structure of the compound.
\item The very high values of the low-temperature specific heat 
measured in these systems are attributed
to the enormous entropy increase associated to the thermal breakdown
of soft, RKKY-induced spin correlations.
\item For higher values of the Kondo coupling, the low-energy electrons
and holes are energetically allowed to participate in the formation
of both uncorrelated and strongly correlated charged modes which
results in a very strong renormalization of the low-energy charged
spectrum. The formation of strongly correlated charged modes is
much more favored in antiferromagnetic Kondo lattices than in
the ferromagnetic ones.
\item Beyond a critical coupling $J_c$, the strongly correlated modes
condensate in the ground state producing an instability and
a transition to a new phase. This transition must be accompanied by 
an important change in the specific heat because the association of
spin waves to conduction electrons to form collective states will
modify the localized spin system.
\end{itemize}

The main future challenges posed by this article are the detailed
study of the spin correlations, its inclusion in the calculation
of the charged modes, and the description and characterization of
the new phase.

\section*{Acknowledgements}
We would like to thank Igor Sandalov for providing discussions 
and very valuable comments. This work has been financed by the DGICYT 
(Research Project No.\ PB93-1249) and by the CIRIT (Grant 1995SGR 00039). 
J.M.~Prats acknowledges financial support from a postdoctoral fellowship 
from the {\em Ministerio de Educacion y Ciencia\/} of Spain.

\vfill

\end{document}